\documentclass[acmsmall]{acmart}
\usepackage{booktabs} 
 \usepackage{enumitem}
\usepackage{balance}
\usepackage[linesnumbered,ruled,vlined]{algorithm2e}
\usepackage{xcolor}
\usepackage{listings}
\usepackage{cleveref}
\usepackage{graphicx}

\setcopyright{none} 
\renewcommand\footnotetextcopyrightpermission[1]{} 
\SetKwInput{KwInput}{Input}                
\SetKwInput{KwOutput}{Output}              

\SetCommentSty{commfont}

\begin{document}
\title{PolyDL: Polyhedral Optimizations for Creation of High Performance DL primitives}


%

\author{Sanket Tavarageri, ~~Alexander Heinecke, ~~Sasikanth Avancha,~~Bharat Kaul}
\affiliation{
  \institution{Intel Labs}
}
\email{sanket.tavarageri@intel.com}

\author{Gagandeep Goyal, ~~Ramakrishna Upadrasta}
\affiliation{
  \institution{IIT Hyderabad}
    }
\email{ramakrishna@iith.ac.in}


\begin{abstract}
Deep Neural Networks (DNNs) have revolutionized many aspects of our lives.
The use of DNNs is becoming ubiquitous including in software for image recognition,
speech recognition, speech synthesis, language translation, to name a few. 
The training of DNN architectures however is computationally expensive. Once the model is created,
its use in the intended application -- the inference task, is computationally heavy too and the inference
needs to be fast for real time use.
For obtaining high performance today, the code of Deep Learning (DL) primitives optimized for specific architectures by expert programmers  exposed via libraries is the norm. However, given the constant emergence of new DNN architectures, creating hand optimized code is expensive, slow and is not scalable. 

To address this performance-productivity challenge, in this paper
we present compiler algorithms to automatically generate high performance implementations of DL primitives that closely match the performance of hand optimized libraries.
We develop novel data reuse analysis algorithms using the polyhedral model
 to derive efficient execution schedules
automatically. In addition, because most DL primitives use some variant of matrix multiplication
at their core, we develop a flexible framework where  it is possible to plug in library implementations of the same in lieu of a subset of the loops. 
We show that such a hybrid compiler plus a minimal library-use approach results in state-of-the-art performance. We develop compiler algorithms to also perform 
operator fusions that reduce  data movement through the memory hierarchy of
the computer system. Using Convolution Neural Network (CNN) models and matrix multiplication operations, we demonstrate
that our approach automatically creates high performing DNN building blocks
whose performance matches the performance of hand-crafted kernels of Intel's oneDNN library on high end CPUs. At the same time, our techniques 
take only a fraction of time ($\frac{1}{20}$ or less) compared to AutoTVM, a deep learning
auto-tuner to create optimized implementations.
\end{abstract}

%
%

%
%

\settopmatter{printfolios=true}
\maketitle

\section{Introduction}
\label{sec:intro}

Deep learning has revolutionized many spheres of human activity, examples of which include, speech recognition \cite{hinton2012deep}, image recognition \cite{krizhevsky2012imagenet,he2016deep}, web search \cite{googleai}, language translation \cite{wu2016google}, conversational artificial intelligence \cite{devlin2018bert} etc. 
Training and inference using deep neural networks (DNNs) that lie at the heart of Deep Learning (DL) are computationally intensive tasks.
 In today's datacenters, predominantly CPUs are  used for inference tasks
partly due to latency considerations.
According to a recent McKinsey study \cite{mckinseystudyinferencehardware},
CPUs account for 75\% of the inference market.
Software frameworks such as TensorFlow, and PyTorch have been 
created to allow data scientists to write high performance deep learning code in 
an efficient manner. 
However, all these frameworks use manually optimized primitives to deliver
high performance.

\begin{figure}[h!]
\begin{minipage}{0.49\textwidth}
\centering
\includegraphics[scale=0.3]{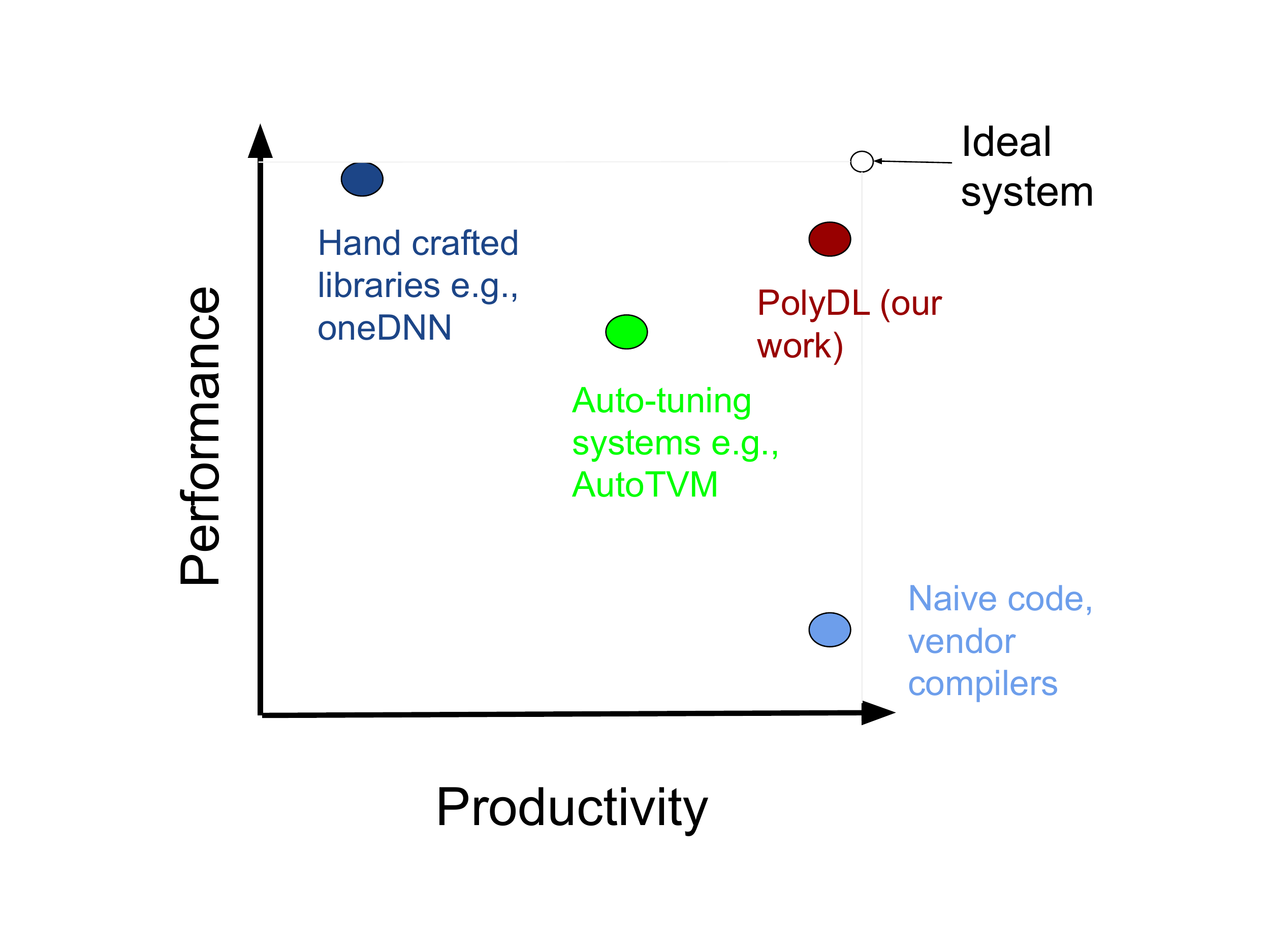}
\caption{The performance, productivity trade-off of different approaches}
\label{fig:perfproductivity}
\end{minipage}
\begin{minipage}{0.49\textwidth}
At the one end of the spectrum, expert coded primitives such as that of Intel oneDNN library attain high performance at the cost of productivity -- expert programmers
have to hand craft the logic of the primitives for the target architectures. Autotuning systems ease the burden on programming to an extent but
do not attain highest levels of performance. Using functionally correct code with vendor supplied compilers is most productive but comes at the expense
of performance. Our work -- PolyDL is close to attaining the highest levels of performance and at the same time being most productive.
\end{minipage}
\end{figure}

Given the ubiquity of CPUs and their widespread use for inference in deep learning applications, in this work we focus on automatically
creating high performance implementations of DL primitives on CPU platforms.
Creating a high performance implementation of a DL primitive requires that the code is parallelized in a load balanced fashion to take
advantage of the multiple cores. Within a single core, the code should be cache friendly: this often means the loops of the code are tiled so that 
effective data reuse
out of different levels of cache (L1, L2, and L3) is possible. Tiling/blocking program transformation \cite{uday08pldi} facilitates data reuse from caches and therefore masks
the long latency of fetching data from the main memory. Additionally, CPUs feature wide SIMD/vector units. Therefore, the loops should 
be adequately vectorized. Oftentimes, the different transformations mentioned are intertwined and that presents challenges for the compiler to
produce fully optimized code automatically.
Existing automatic compilation, and auto-tuning techniques  \cite{uday08pldi,Kong:2019:MTM:3314221.3314653,tavarageri2013adaptive,renganarayanan2007parameterized,
darte2014parametric,baskaran2010parameterized,hartono2009parametric,
tavarageri2010parametric,chen2018learning,baghdadi2019tiramisu,chung2004using,
chen2007model,tiwari2009scalable} are either 1) inadequate to generate code that matches
the performance of hand-tuned library implementations -- later on in the paper we show
that the state-of-the-art compiler generated code can lag library implementations
by as much as \textasciitilde 10X or more, or 2) expensive -- it would require running of 1000s of code versions to discover the best performing version and yet, fall short of 
reaching the peak performance of the machine. 
The main reason for the failure of automatic compilation techniques in achieving very high performance levels needed is that
the sophistication in the CPU microarchitecture has increased over successive generations of CPUs (data prefetching, speculative execution, vector units, deep memory hierarchies, complex cache replacement algorithms etc). Consequently, 
the cost functions used to optimize code are unable to capture the nitty-gritties of the underlying architectures, and therefore are unable to derive the most effective execution schedules.
Auto-tuning is an alternative approach where one explores a large number of
program variants and selects the best performing version, sidestepping the 
complexity of defining a cost function that adequately models the 
intricacies of the underlying hardware architecture. However, auto-tuning is
expensive and furthermore, it may fall short of manually created library in performance as our study shows later on in the paper with respect to AutoTVM \cite{chen2018learning}, 
a deep learning auto-tuning system.
We characterize the performance, productivity trade-off qualitatively in Figure \ref{fig:perfproductivity}.

It has been shown that 95\% of all deep learning applications running in the
data centers today have a recurring pattern in their inner most loops, namely blocked
matrix multiplication \cite{jouppi2017datacenter,georganas2020harnessing}.
We decompose the overall DL primitive optimization problem into two parts:
1) efficient parallelization of the code and discovering a structure of the loops that uses the multi-level caches well, and
2) effective vectorization of the code for a high degree utilization of the vector units.
In this paper, we develop a novel polyhedral model based data reuse algorithm to derive load balanced parallel loops and cache friendly execution schedules.
For the latter i.e., to use the vector units optimally, we develop a flexible framework where
the inner most loops of kernels can be replaced with \emph{microkernels}, i.e., manually optimized library implementations.
As the number of recurring patterns in the inner most loops of DL primitives is small, our hybrid approach is more scalable because
the number of microkernels that the expert programmers have to create is small as well.
Thus, the problem of DL library development will now be reduced to hand coding a few microkernels as opposed to hand coding
each of the large number of DL primitives.

To account for the sophisticated memory hierarchy, we use a code-generator to create a number of program variants - $n$ in number -  for a given program. 
The generated code variants are then analyzed by our novel data reuse algorithm -- 
\emph{PolyDL} to characterize their cache behavior.
We have developed a \emph{relative ranking} algorithm which ranks the $n$ variants based on their potential performance. The top $k$ variants are selected and are run on the target hardware
and the best performing program version is discovered.
Thus, \emph{PolyDL} narrows down the number of variants to actually run on the target architecture
from $n$ to a small, manageable $k$ ($k << n$). 
Through our experimental evaluation on convolutions of a range of popular and the state-of-the-art image recognition models, we show that
the top variant (a single variant) picked by our compilation machinery is one of the best performing variants, and
the realized performance is close to and in many cases, higher than that of Intel's oneDNN library \cite{intelmkldnn} (formerly known as MKL-DNN), a hand-tuned library for deep learning kernels. Additionally, we develop a fusion algorithm that ``fuses'' element-wise operators with their preceding or succeeding compute-intensive operators. Such patterns where a computationally heavy operator such as convolution is succeeded by element-wise operators such as \textsf{ReLU}, occur frequently in DL workloads. Therefore, the DL domain specific  fusion algorithm that we develop will increase the overall performance by reducing the extra memory traffic that the element-wise operators would otherwise incur.

The contributions of the paper are the following:
\begin{itemize}
  \item We present a novel cache data reuse analysis to characterize a loop nest's behavior with respect to a multi-level cache hierarchy.
   \item We describe a methodology to rank program variants in terms of performance using the compiler generated
   statistics and the system parameters, i.e., cache sizes. 
   To this purpose, we develop two ranking techniques: one, a heuristic for ranking and two, a DNN based approach.
   \item We develop a deep learning domain specific operator fusion algorithm.
   \item We conduct extensive experiments comparing our technology with the Intel oneDNN library and with AutoTVM.
   The experiments show that we are able to match the performance of expert coded DL primitives in the oneDNN library and 
   exceed the performance of the ones discovered by AutoTVM via extensive auto-tuning.
\end{itemize}

To the best of our knowledge, this is the first work that examines
  automatic compilation techniques and the use of microkernels  in an integrated
  fashion.
The rest of the paper is organized as follows.
We motivate the need for derivation of automatic execution schedules for loops in Section \ref{sec:motivation}.
Section \ref{sec:background} describes preliminary concepts that will 
be used in developing the compiler algorithms. 
In Section \ref{sec:reuse}, we develop algorithms for compile-time selection of top performing code version(s). We present a cache data reuse analysis and a poly-ranking
system to rank the candidate program variants in terms of performance.
The operator fusion algorithm is expounded in Section \ref{sec:fusion}.
Section \ref{sec:experiments} details the experimental evaluation conducted.
The related work is discussed in Section \ref{sec:related}
while Section \ref{sec:conclusion} presents the conclusions from this work.

\section{Motivation}
\label{sec:motivation}
\begin{figure*}[h!]
\centering
\begin{minipage}{0.49\textwidth}
\centering
\includegraphics[scale=0.37]{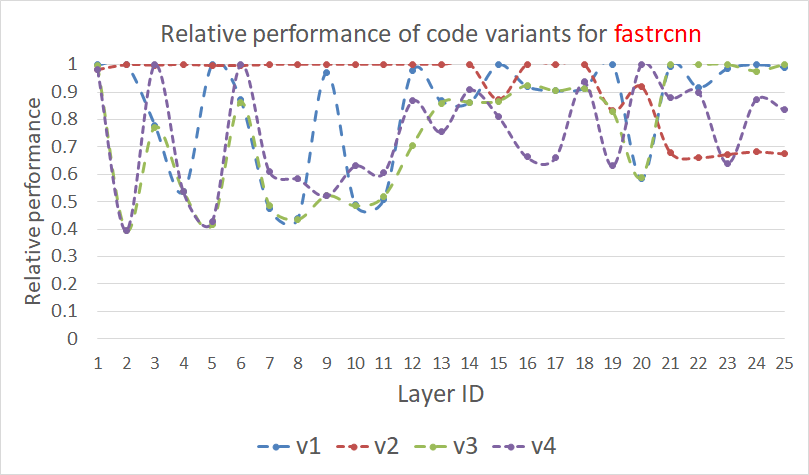}
\caption{Performance of four code variants}
\label{fig:fastrcnn_motivation}
\end{minipage}
\begin{minipage}{0.49\textwidth}
\centering
\includegraphics[scale=0.37]{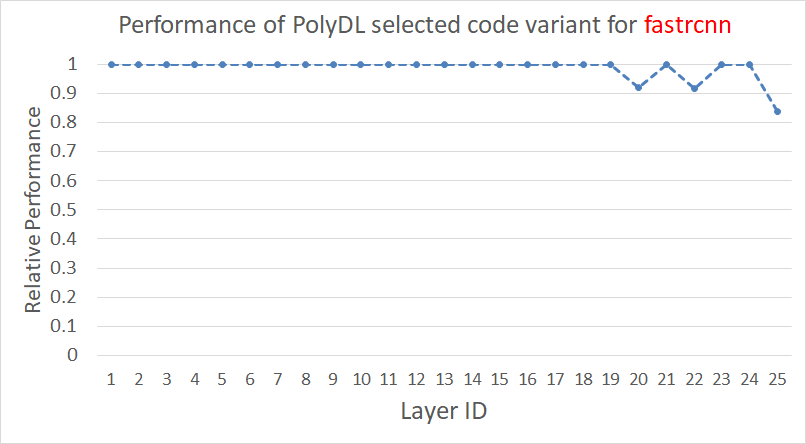}
\caption{Performance of the PolyDL picked variant}
\label{fig:fastrcnn_motivation_polyscientist}
\end{minipage}
\end{figure*}

The deep learning primitives are computationally intensive and most of the 
neural network training and inferencing time is spent in them.
However, for different layers of the deep neural networks, the optimizations
 (e.g., loop order, tile sizes in tiled code etc) that need to be applied are different.
 Using a version of the code optimized for one layer of a neural network for all
 others can yield poor performance for the overall neural network.
 It is this need for custom optimization for different layers of neural networks (the number of layers in a deep neural network can be large) that makes generating efficient code for
 deep learning primitives a challenging problem.
 To illustrate the need for such tailored loop optimizations we consider 
 the convolution layers of the Fast R-CNN model  \cite{girshick2015fast},
 one of the leading image recognition CNN models.
 We generate four variants of convolution code which differ only in the loop order and the rest of the loop structure remains the same for all of them (more details are
 provided in \S \ref{sec:experiments}) and measure performance on 
 a 28-core Intel(R) Xeon(R) Platinum 8280 (a.k.a Cascade Lake) CPU server.
 Figure \ref{fig:fastrcnn_motivation} shows the normalized performance
 of the code variants on 25 convolution layers of Fast R-CNN: 
 the performance is normalized with respect to the highest performing code
 among the four variants.

From Figure \ref{fig:fastrcnn_motivation}, we observe that the performances of 
different versions of the code vary widely from layer to layer. 
A convolution layer differs from another convolution layer in problem sizes -- 
image sizes, channel widths,
filter sizes, strides, and padding.
The code version \emph{v2} has the best performance from layer 1 through 19,
and is subpar from layer 20 through 25. 
The efficiencies of the other versions viz., \emph{v1}, \emph{v3}, and \emph{v4}
are much more widely varying.
Using the compiler technology
we have developed --  \textsf{PolyDL} that we detail in the rest of the paper, we are able to effectively analyze
the four code variants and pick the best performing variant for each layer.
The performance achieved by PolyDL picked code shown in Figure \ref{fig:fastrcnn_motivation_polyscientist} is close to the highest
performance among the four variants for all 25 layers of Fast R-CNN.
Thus using the PolyDL system, using compile-time static analysis alone, we are able to automatically identify and apply
the loop optimizations required for each layer of a deep neural network in order to 
achieve high performance.

\section{Preliminaries}
\label{sec:background}

\subsection{Notation}
We use the polyhedral model \cite{feautrier1996automatic}, which is an advanced mathematical framework to reason about dependences and loop transformations, to develop our data reuse algorithm. 
We use the Integer Set Library \cite{verdoolaege2010isl} for performing polyhedral operations in this work and we use the same notation as used in ISL to elucidate the concepts and the algorithm.
The matrix multiplication code shown in Figure \ref{matmulcode} will be used to illustrate the workings of the data reuse analysis.

\begin{minipage}{.49\textwidth}
\begin{figure}[H]
\begin{lstlisting}[language=C,basicstyle=\scriptsize,frame=bottomline]
for (i = 0; i < M; i++) {
    for (j = 0; j < N; j++) {
	    for (k = 0; k < K; k++) {
    	    C[i][j] += A[i][k] * B[k][j];
        }
    }
}
\end{lstlisting}
\caption{Matrix multiplication code}
\label{matmulcode}
\end{figure}
\end{minipage}
\begin{minipage}{.49\textwidth}
\paragraph{Sets}
A set is a tuple of variables $x_i$s along with a collection of constraints $c_k$s defined on the tuple variables. 
$ s = \{ [x_1, \dots , x_n] : c_1 \land \dots c_m \} $

The iteration spaces of loop nests are represented by sets. The iteration space of the loop in Figure \ref{matmulcode} is defined as the following set.
$ I = \{ S[i, j, k] : 0 <= i < M \land 0 <= j < N \land 0 <= k < K \} $
\end{minipage}

\paragraph{Relations}
A relation is a mapping from input tuple variables $x_i$s to output tuple variables $y_j$s.
In addition, a set of constraints $c_k$s can be defined for a relation that will place constraints on the input/output tuple variables. 
$ r = \{ [x_1, \dots, x_n] \mapsto [y_1, \dots, y_m] :  c_1, \dots, c_p \}$

The read and write access functions of a loop nest can be modeled with relations. The read relations in the Figure \ref{matmulcode} code are shown below: 
$r_1  = \{ S[i, j, k] \mapsto  C[i, j] \}$,
$r_2  = \{ S[i, j, k] \mapsto  A[i, k] \}$,
$r_3  = \{ S[i, j, k] \mapsto  B[k, j] \}$.
The sole write relation in the loop is: $w_1 = S[i, j, k] \mapsto C[i, j]$.
The domain of a relation $r$ is denoted by {\textsf dom} $r$.

\paragraph{Apply operation} When a relation $r$ is applied on a set $s$, the domain of $r$ will be intersected with $s$ and the resulting range will be a new set $s'$. The set $s'$ is said to be the result of the apply operation. The operation is mathematically defined as:
$ ( \vec{y} \in s') \Longleftrightarrow (\exists \vec{x} ~~\text{s.t}~~ (\vec{x} \in s \land \vec{x} \mapsto \vec{y}) \in r )$

The data footprint of the loop can be computed by \emph{applying} read and write \emph{relations} on the iteration space \emph{set}:
$ r_1(I) \cup r_2(I) \cup r_3(I) \cup w_1(I) $

\paragraph{Lexicographic operations} The lexicographical operations can be applied on
 sets. $s_1 << s_2$ outputs all the elements of $s_1$ that are  lexicographically 
strictly smaller than all the elements of $s_2$, while $s_1 <<= s_2$ gets us the elements of $s_1$ that are lexicographically smaller than or equal to the elements of $s_2$.
The lexicographically smallest element of a set $s$ is queried using {\textsf lexmin} $s$.
Similarly, the lexicographically largest element is obtained using {\textsf lexmax} $s$.

\paragraph{Set difference.} The set difference
between set $s_1$ and $s_2$ is denoted by $s_1 - s_2$, i.e., the resulting set will have
elements of $s_1$ that do not appear in $s_2$.

\subsection{Polyhedral dependences}
The exact data dependences in loop nests can be computed in the polyhedral model and are expressed as maps from source iterations to target iterations involved in the dependence. For cache data reuse analysis developed in \S\ref{sec:reuse}, we consider  four kinds of dependences -- Read-After-Read (RAR), Read-After-Write (RAW, a.k.a \emph{flow}), Write-After-Read (WAR, a.k.a \emph{anti}), and Write-After-Write (WAW). 
  The data dependencies of the matrix multiplication code in Figure \ref{matmulcode} are shown below.
  
\begin{align*}
d_1 = & \{ S[i, j, k] \mapsto S[i', j', k'] : i' = i \land j' = j \land k < k' < K \} \\
d_2 = & \{ S[i, j, k] \mapsto S[i', j', k'] : i' = i \land k' = k \land j < j' < N \} \\
d_3 = & \{ S[i, j, k] \mapsto S[i', j', k'] : j' = j \land k' = k \land  i < i' < M \} \\
\end{align*}

The dependence $d_2$ is induced by array reference A[i][k]. An element of array A, say A[0][0] which is accessed in \emph{source} iteration $[i=0,j=0,k=0]$ gets reused in \emph{target} iterations $[i'=0,j'>0,k'=0]$. The source to target iteration relationships such as this are expressed in a parametric fashion as the relation $d_2$. 

\begin{figure*}[t!]
\centering
\includegraphics[scale=0.4]{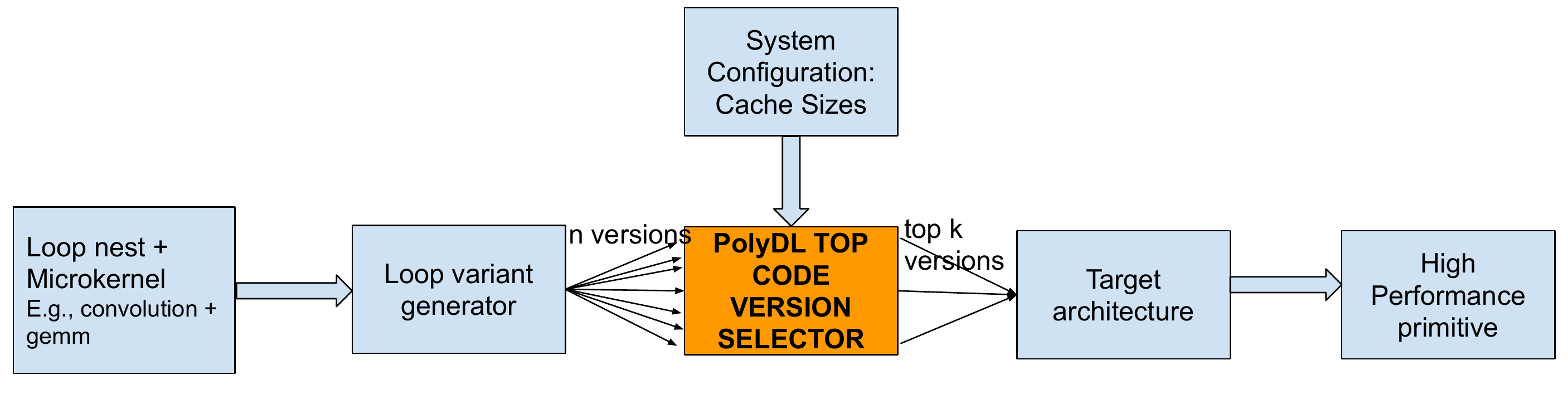}
\caption{The PolyDL system}
\label{fig:system}
\end{figure*}

\section{Compile Time Selection of Top performing code version}
\label{sec:reuse}

The input to our compiler tool is,
the loop nest to be optimized -- $\mathcal{L}$ along with the microkernel
 that forms the inner-most loops.
Figure \ref{fig:system} shows the overall system design. 
The loop based specification of the microkernel -- $\mathcal{M}$ is substituted in the code for further analysis. The resulting loop structure -- $\mathcal{L'}$ is regular.
The code generator takes the loop nest $\mathcal{L'}$ and generates a large number
of program variants while keeping the inner most loops that correspond to 
$\mathcal{M}$ intact.
For each generated code variant, the working set sizes are computed as described in 
\S\ref{sec:wscompute}.
The statistics calculated for all the variants are then input to the poly-ranking algorithm
described in \S\ref{sec:wholepolyranking} and it picks the top $k$ best performing versions.
The original microkernels are inserted back into the code of the $k$ picks.
The top performing variants selected analytically are now run on the target
architecture and 
 best performing code among the $k$ loop nests is determined.

\textbf{Microkernel Specification}
The microkernel function call is annotated with a pragma compiler directive which contains the loop-based functionally equivalent code.
The microkernel function call is substituted with the loop based code for the compiler
analysis in a pre-processing pass. When the cache data reuse analysis and ranking of
the code variants are done, in a post-processing pass, the loop-based inner most loops
are replaced with the call to the microkernel.

We assume that the data accessed by the microkernel's equivalent loop based code and by the implementation of the microkernel are the same.
The order of the loops within the microkernel implementation could be potentially different.
There is no assumption on how the microkernel is implemented. The only assumption is that the data set accessed should be the same. 

\begin{algorithm}  
\scriptsize
 \KwInput{Loop nest: $\mathcal{L}$}
 \KwOutput{The working set sizes: $WS_{\text{all}}$}
\{Iteration space: $\mathcal{I}$, Read relations: $r_{\text{read}}$, Write relations: $r_{\text{write}}$, Schedule: $\delta$\} $\leftarrow$ Parse the loop nest $\mathcal{L}$ \\
\{$\mathcal{D}_{\text{RAR}}, \mathcal{D}_{\text{RAW}}, \mathcal{D}_{\text{WAR}}, \mathcal{D}_{\text{WAW}}$\} $\leftarrow$ Compute read-after-read, read-after-write, write-after-read, write-after-write dependences of $\mathcal{L}$ \\
$\mathcal{D}_{\text{all}} \leftarrow \mathcal{D}_{\text{RAR}} \cup \mathcal{D}_{\text{RAW}} \cup \mathcal{D}_{\text{WAR}} \cup \mathcal{D}_{\text{WAW}}$ \\
$WS_{\text{all}} \leftarrow \emptyset$ \\
\tcc{Iterate through all dependences to compute the working set sizes}
\For{$d \in \mathcal{D}_{\text{all}}$}{
  \If{$d$ spans parallel iterations}{
   \tcc{The dependence spans iterations of the parallel loop(s)}
     $i_p \leftarrow$ The outermost parallel iterator in dom $d$ \\
     $\mathcal{I}_{\text{par}} \leftarrow$ Parameterize iterators outer to $i_p$ in $\mathcal{I}$ \\
     $WS_{\text{par}} \leftarrow | r_{\text{read}}(\mathcal{I}_{\text{par}}) \cup r_{\text{write}}(\mathcal{I}_{\text{par}})|$ \\
     Add $WS_{\text{par}}$ to $WS_{\text{all}}$
     }\Else {
     \tcc{The dependence spans iterations of the sequential loops}
 $\mathcal{I}_{\text{source}} \leftarrow \text{lexmin dom}~~~ d$ \\
 $\mathcal{I}_\text{{min\_tar}} \leftarrow \text{lexmin}~~~ d(\mathcal{I}_\text{source})$ \\
 $\mathcal{I}_\text{max\_tar} \leftarrow \text{lexmax}~~~ d(\mathcal{I}_\text{source})$ \\
 $\mathcal{I}_\text{min\_WS} \leftarrow (\mathcal{I} <<= \mathcal{I}_\text{min\_tar}) 
 -  (\mathcal{I} << \mathcal{I}_\text{source})$\\
  $\mathcal{I}_\text{max\_WS} \leftarrow (\mathcal{I} <<= \mathcal{I}_\text{max\_tar}) 
 -  (\mathcal{I} << \mathcal{I}_\text{source})$\\
 $WS_\text{min} \leftarrow | r_\text{read}(\mathcal{I}_\text{min\_WS}) \cup r_\text{write}(\mathcal{I}_\text{min\_WS})|$ \\
  $WS_\text{max} \leftarrow | r_\text{read}(\mathcal{I}_\text{max\_WS}) \cup r_\text{write}(\mathcal{I}_\text{max\_WS})|$ \\
  Add $WS_\text{min}$ and $WS_\text{max}$ to $WS_\text{all}$
}
}
\caption{Compute working set sizes}
\label{alg:wscompute}
\end{algorithm}

\subsection{Working set size computation}
\label{sec:wscompute}

We develop a polyhedral model based cache data reuse analysis to characterize a loop-nest's behavior with respect to a given cache hierarchy. The analysis computes the various existing data reuses of a program and then for the input cache hierarchy determines which data reuses are exploitable at various levels of cache.

Each data dependence in a loop is also a case of data reuse -- the source and target iterations involved in the dependence touch the same data element and therefore, the data is reused. For a data dependence and hence data reuse to be realizable in a given level of cache, all the data elements accessed between the source and target iterations of the dependence -- the \emph{working set} -- have to be retained in the cache so that when the execution reaches the target iteration, the data element(s) used in the source iteration will still be present in the cache.

Algorithm \ref{alg:wscompute} computes all the working sets of the input loop nest.
First, the input C source file is parsed using the Polyhedral Extraction Tool (PET) \cite{Verdoolaege2012pet} to obtain the polyhedral representation of the program, namely iteration space of the loop nest, read and write relations and the schedule (line 1).  The exact (and not transitive) RAR, RAW, WAR, WAW dependences are then computed and a union of all the four kinds of dependences is formed (line 2 and 3). 
The task now is to compute the working set size for each dependence which is carried out from line 6 through 19. 

We distinguish between the data dependences that span parallel iterations and those that do not. The working set sizes for the two kinds of dependences are computed differently.
If a dependence spans a parallel loop and the data set used in the entire set of parallel iterations is held in the given level of cache, then the data reuse is guaranteed to happen out of that cache. This is because, irrespective of the order of execution of the iterations of the parallel loop, the data accessed by the source and target iterations will be present in the cache as the cache is big enough to hold the entire set of data elements accessed by all parallel iterations collectively. The working set for such a dependence is calculated by parameterizing the iterations outer to the parallel loop variable and evaluating the size of the read and write sets within the parameterized iteration set (line 7 to 9).

For the data dependences that span sequential loops, we compute the working set size as follows. We consider a representative \emph{source} -- the first iteration (lexicographically) of all the source iterations of a dependence (line 12).
We can now compute the target iterations for the lexicographically first/minimum iteration. If the data element that is used in the source iteration is used in multiple subsequent iterations then there may be multiple target iterations for the same source iteration. Therefore, the working sets to exploit the data reuse may vary. 
For this reason, we compute the first (i.e., lexicographically minimum) and the last (i.e., lexicographically maximum) iterations of the target iteration set (line 13 and 14).
The intervening iterations between the source and the first target iteration are determined (line 15). Similarly, iterations between the source and the last target iteration are derived (line 16).
The working sets will be the union of all the read and written data elements between 
the source and the first/last iterations of the target iteration set  (line 17 and 18).
Correspondingly, for each dependence we compute two working set sizes -- $WS_{min}$ and $WS_{max}$, if there are multiple target iterations for a source iteration in a given dependence.
What this means is, in order to be able to exploit at least one data reuse arising from the dependence $d$, the cache memory should be capacious enough to hold at least $WS_{min}$ data elements. If all the data reuses are to be realized -- till the last target iteration, then the cache should of size equal to or greater than $WS_{max}$ times the datatype size.

We illustrate the operation of the algorithm using the running example in Figure \ref{matmulcode}. Let us examine the following dependence carried by the $j$ loop arising because of the array reference $A[i][k]$:
$d_2 =  \{ S[i, j, k] \mapsto S[i', j', k'] : i' = i \land k' = k \land j < j' < N \}$.
Of all the source iterations, the first/lexicographically minimum iteration is: 
$\mathcal{I}_{source} = \{ S[i = 0, j = 0, k = 0] \}$
Its target iterations are: 
$\{ S[i = 0, j, k = 0] :  0 < j < N \}$.
Among the target iterations, the first one is: 
$I_{min\_tar} = \{ S[i = 0, j = 1, k = 0]  \}$ 
and the last one is: 
$I_{max\_tar} = \{ S_3[i = 0, j = N-1, k = 0]  \}$

The number of data elements of the three arrays -- A, B, C accessed between $\mathcal{I}_{source}$ and $I_{min\_tar}$ is derived by \emph{applying} the read and write relations on the intervening iteration set and it is: $$WS_{min} = 2K + 3$$

The $K$ elements of array A -- $A[0][0, 1, \dots, K-1]$, the $K+1$ elements of array B --
$B[0, 1, \dots, K-1][0]$ and $B[0][1]$, and finally $2$ elements of array C -- 
$C[0][0], C[0][1]$ accessed between the source iteration $S[i = 0, j = 0, k = 0]$
and the target iteration $I_{min\_tar} = S[i = 0, j = 1, k = 0]$ lead to the $WS_{min}$ size of $2K + 3$.

The maximum working set size -- the size of the data touched between $\mathcal{I}_{source}$ and $I_{max\_tar}$ is:

$$ WS_{max} = N\times K + N +1 $$
The $WS_{max}$ size is arrived at by counting the number of array elements
accessed between the source iteration - $S[i = 0, j = 0, k = 0]$ and the target iteration - 
$I_{max\_tar} = \{ S_3[i = 0, j = N-1, k = 0]  \}$.
As far as array A is concerned, $K$ elements of it -- $A[0][0, 1, \dots, K-1]$ are read.
Array B's elements -- $B[0, 1, \dots, K-1][0, 1, \dots, N-2]$ plus $B[0][N-1]$ are read which total $K \times (N-1) + 1$.
$N$ elements of array C are read and written -- $C[0][0, 1, \dots, N-1]$. Therefore, a total of $N\times K + N +1$ are read and written.

\subsection{Poly-ranking algorithm}
\label{sec:wholepolyranking}

\begin{algorithm} 
\scriptsize
 \KwInput{The working set sizes: $WS_{all}$, \\
  Cache sizes: $Size_{L_1}, \dots Size_{L_n}$} 
 \KwOutput{Working set sizes per cache: $WS^{L_i} ~~\text{for}~~ i = 1, \dots, n$, \\
 Memory working set size: $WS^{mem}$}

Initialize $WS^{L_i} ~~\text{to} ~~ 0~~\text{for}~~ i = 1, \dots, n$, \\
Sort working set sizes in $WS_{all}$ from smallest to largest \\
\For{$WS_j \in WS_{all}$}{
	\For{$Size_{L_i} \in Size_{L_1}, \dots Size_{L_n}$}{
		\If{$(WS_j + WS^{L_i}) \le Size_{L_i}$} {
			$WS^{L_i} = WS^{L_i} + WS_j$ \\
			\textbf{break} \\
		}
	}
}

Add the working sets $WS_j \in WS_{all}$ that do not fit any cache to $WS^{mem}$
\caption{Compute working set sizes w.r.t cache sizes}
\label{alg:cachewscompute}
\end{algorithm}

We have built a code generator to emit a number of program variants.
The code generator creates the loop variants by applying tiling and loop interchange
program transformations. The tile sizes are varied as well.
The working set size computation analysis --\S\ref{sec:wscompute} is performed on each program version generated.
Among the many variants generated, the poly-ranking algorithm described below picks
the top $k$ best performing versions, where $k$ is a parameter.

We assume fully associative, and exclusive caches. 
If the working set size corresponding to a data reuse in the program is smaller
than the cache size then the data reuse is exploitable in the cache.
The poly-ranking system considers caches at different levels (typically L1, L2, and L3)
and for each data reuse, determines at what level of cache hierarchy is the data reuse
realizable.
Algorithm \ref{alg:cachewscompute} shows the steps to determine the cumulative
working set sizes at each level of cache. The inputs to the algorithm are the 
working set sizes computed for a loop nest, and the cache sizes of the target system.
The algorithm determines the fastest level of cache where the working set size corresponding to each data reuse fits
and adds it to that cache's working set size. 
The working set sizes that fit in a particular level of cache $L_i$ are denoted by $WS^{L_i}$.
If a working set does not fit in any cache, then the data reuse happens
out of the main memory. Consequently, the memory's working set size is updated.

\subsubsection{Performance cost model based ranking}
\label{sec:polyranking}
The running time of the loop is directly related to the latency of the cache where the data
reuse occurs as well as the working set size. Furthermore, the running time is inversely
related to the bandwidth of the cache.
Based on these observations, we define the following cost function:

\begin{align}
\mathcal{C} = & \sum_{L_i} WS^{L_i} \times \frac{\text{lat}^{L_i}}{\text{bw}^{L_i}}  + WS^{\text{mem}} \times \frac{\text{lat}^{\text{mem}}}{\text{bw}^{\text{mem}}} \label{eq1} 
\end{align}

The latency of cache $L_i$ is $\text{lat}^{L_i}$ while its bandwidth 
is denoted by $\text{bw}^{L_i}$.
For each code variant generated, we run the cache data reuse analysis and
calculate the above cost function. 
Then, the variants are ranked in the decreasing order of the value of the cost function.
The working set size at cache level is multiplied with that cache's latency and divided by its bandwidth. The multiplicands are then added together.
The lower the value of the cost function, the higher is its presumed performance, and higher is its rank.

\subsubsection{DNN-based ranking algorithm}
\label{sec:dnnranking}

We explore the use of deep neural networks (DNNs) for 
ranking of code variants. 
For the purposes of training the DNN model, 
we collect the performance data of code variants generated and
the statistics as outputted by Algorithm \ref{alg:cachewscompute} -- 
working set sizes at different levels of the memory hierarchy.

We train the DNN model to perform relative ordering of \emph{two} code
variants.
We then use a \emph{tournament} based ranking system to assign ranks
to the different code versions created -- 
we play each code variant against every other code variant.
For each variant, we record the number of wins it has accumulated.
We then rank the variants based on the number of wins -- 
the higher the number of wins, the higher the rank.

\begin{minipage}{0.6\textwidth}
\begin{figure}[H]
\centering
\includegraphics[scale=0.45]{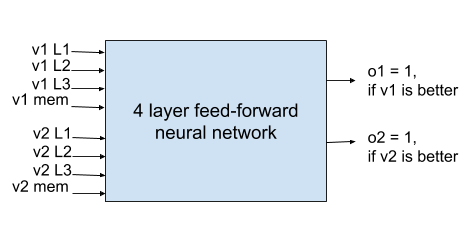}
\caption{The DNN architecture for ranking of code variants}
\label{fig:polydnn}
\end{figure}
\end{minipage}
\begin{minipage}{0.4\textwidth}
We use four intermediate layers of 64, 32, 16, 8 neurons respectively.
We use \emph{relu}, \emph{relu}, \emph{softsign}, and \emph{relu}
activation functions for the four intermediate layers.
\end{minipage}

We use a four layer feed forward neural network architecture shown in Figure \ref{fig:polydnn}.
We normalize the compiler generated statistics of two code variants 
in the following fashion and input them to the DNN.
We sum the working set sizes of the two variants together:
$sum = WS^{L_1}_{v_1} + WS^{L_2}_{v_1} + WS^{L_3}_{v_1} + WS^{mem}_{v_1} 
+ WS^{L_1}_{v_2} + WS^{L_2}_{v_2} + WS^{L_3}_{v_2} + WS^{mem}_{v_2}$
and divide the individual statistic by this sum.
The rationale for considering the sum of the two statistics together is that
if one of the variants is creating higher volume working set sizes then
its statistics should appear bigger to the DNN.
This is because the smaller the working set sizes, we can expect higher
performance.
Therefore, for the DNN to learn the relative performances of the two variants,
it is crucial that it sees the relative sizes of the working set sizes.
Normalizing each variant individually (by considering the sum of statistics of one
variant alone) would not bring out the differences in the absolute values
of the working set sizes of the two variants at different cache levels.
The output layer consists of two neurons and we use the \emph{softmax}
function for the output layer.
The values of the two output neurons, because of the use of the softmax function,
sum to 1. 
If the output value is above a threshold - $\theta$, we consider it a 1, otherwise a 0.
If the first neuron fires a 1, then the first variant is considered the winner.
If the second neuron fires a 1, then the second variant is considered the winner.
If both of them are zero because none of them are above the threshold, then
it is a draw between the two variants. In this work, we set the threshold $\theta$ to 0.6.
We experimented with deeper models as well. However, the depth beyond 
four layers did not have any discernible effect on accuracy.

\section{Operator Fusion}
\label{sec:fusion}
Often, in the DNN architectures, a \emph{heavy} operator (where most of the compute 
cycles are spent) is followed by activation functions which are \emph{element-wise} operators.
The output of the heavy operator is processed by the activation functions such as {\textsf ReLU, Sigmoid etc.} in an element-wise fashion, i.e., without involving any reduction.
The element-wise operator is mainly a memory bound operator.
fused with the heavy operator in order that the extra data movement through the memory hierarchy
that would otherwise be necessitated by the element-wise operator is eliminated.

\begin{minipage}{0.6\textwidth}
\begin{algorithm}[H] 
\scriptsize
 \KwInput{Loops of computationally heavy operator:  $op_{hy}$, \\
  loops of element-wise operator $op_{ew}$} 
 \KwOutput{Fused operator: $op_{fused}$}

$\mathcal{W}_{hy} \leftarrow $ the write set of $op_{hy}$ \\
$\mathcal{W}_{ew} \leftarrow $ the write set of $op_{ew}$ \\

 $op_{fused} \leftarrow \emptyset$

\If {$\mathcal{W}_{hy} = \mathcal{W}_{ew}$}{
 \If {$|I^{op_{ew}}| =  |\mathcal{W}_{ew}|$}{
  \If {No writes or reads to any element of $\mathcal{W}_{hy}$ between $op_{hy}$ and $op_{ew}$} {
  // We will now fuse the two ops \\
     $\mathcal{I}_{ew} \leftarrow$ instructions in the inner most loops of $op_{ew}$ \\
     $op_{fused} \leftarrow$ Insert $\mathcal{I}_{ew}$ in the last iteration of $op_{hy}$'s reduction loops\\
     $op_{fused} \leftarrow$ Apply index set splitting on $op_{fused}$
  }
  } 
}

\If{$op_{fused} = \emptyset$}{
  // We will return the original loop nests \\
  $op_{fused} \leftarrow \{ op_{hy},~~ op_{ew}\}$
}

\caption{Perform operator fusion}
\label{alg:fusion}
\end{algorithm}
\end{minipage}
\begin{minipage}{0.4\textwidth}
Algorithm \ref{alg:fusion} presents a generic algorithm that fuses a heavy operator with the subsequent element-wise operator. 
\end{minipage}
The following conditions have to be met for an element-wise operator
$op_{ew}$ to be fused with the arithmetically intensive, heavy operator $op_{hy}$:

\begin{itemize}
 \item The two operators --- $op_{ew}$ and $op_{hy}$ should be writing to the same set of elements (line 4 in the algorithm)
 \item The element wise operator $op_{ew}$ should be writing to each array element
   only once. We check if the cardinality of the iteration space of $op_{ew}$ is equal to
   the cardinality of the write set of $op_{ew}$ (line 5). This check will confirm that
   $op_{ew}$ is indeed an element-wise operator and does not involve a reduction.
   \item  The operator $op_{hy}$ should be immediately followed by $op_{ew}$ 
   without any intervening code. Or, if there is any code between the two operators,
   it should not be writing to or reading from the write set of the two operators (line 6).
\end{itemize}
Once the aforementioned conditions are met, the instructions of $op_{ew}$
are inserted in the last iteration of $op_{hy}$ operator's reduction
loops subsequent to the instructions of $op_{hy}$ (line 9). To reduce the overheads stemming from the conditional checking if 
an iteration of the $op_{hy}$ operator's loops is the last iteration of the reduction loops, we apply
index set splitting transformation to peel the last iteration  from the rest of the iterations. 

A symmetric analysis can be applied to fuse an element-wise
operator with a subsequent heavy operator as well. In that case the operation of the
element-wise operator will be fused with the first iteration of the heavy operator's reduction loops.

\section{Experimental Evaluation}
\label{sec:experiments}
We conduct experiments to evaluate the efficacy of the PolyDL system 
in its ability 1)  to derive high performance primitive implementations, and 2) to create
efficient fused operators.
PolyDL's goals are two-fold: one, to achieve performance competitive with manually optimized code,
and two, to do so with compile-time analyses alone without requiring auto-tuning which can be
expensive.
Accordingly, 
we gauge the performance of PolyDL against a state-of-the-art library created
specifically for deep learning networks -- the latest version of Intel oneDNN \cite{intelmkldnn} viz., v1.4.
We also compare PolyDL's performance with AutoTVM system's.
AutoTVM is an auto-tuning system -- it generates a large number  of program
variants, runs them on the target architecture, and observing the performance of different variants, identifies the best performing variant.

\subsection{Set up}
In this work,
we evaluate the performance benefits of the PolyDL system on CPUs for inference tasks.
The forward pass convolution operation, batch-normalization, and ReLU and its variants 
form the bulk of the compute of inference CNN models for image recognition.
The GEMM operation is at the heart of Fully Connected (FC) layers
and multi-layer perceptron (MLP) implementations. 
Therefore, we focus on them in our experimental evaluation.

The experiments are run on the latest Intel servers -- Intel(R) Xeon(R) Platinum 8280 (Cascade Lake) CPU servers running at 
the frequency of 2.70 GHz.
A single socket processor has 28 cores, 32KB private L1 cache, 1MB private L2 cache, and
39MB shared L3 cache.
The programs are compiled with Intel icc compiler 19.0.3.199 with the highest optimization flag -O3.

\subsection{The GEMM microkernel}
\label{subsection:GEMMmicrokernel}
We use the LIBXSMM \cite{libxsmm,heinecke2016libxsmm,georganas2020harnessing}
implementation of GEMM (GEneral Matrix Multiplication) as
the \emph{microkernel}.
\textbf{The data used by a microkernel fits in the registers or at most is L1 cache resident.}
The microkernel performs the computation: $C = \beta . C + \alpha . A . B$, where A, B, and C are matrices and $\alpha$ and $\beta$ are scalars.

\begin{minipage}{.49\textwidth}
\begin{algorithm}[H]
 \KwInput{$A \in \mathbb{R}^{m \times k}, B \in \mathbb{R}^{k \times n}, C \in \mathbb{R}^{m \times n}, \alpha, \beta \in \mathbb{R}$} 
 \KwOutput{$C = \beta . C + \alpha . A . B$}
 
 acc\_regs $\leftarrow$ load C \\
\For{$i_k = 0$ $\dots$ k-1 with step 1}{
   // Perform outer product \\
   acc\_regs += A column$_{i_k}$ $\times$ B row$_{i_k}$.
}

C $\leftarrow$ acc\_regs

\caption{The GEMM Microkernel}
\label{alg:gemmmicro}
\end{algorithm}
\end{minipage}
\begin{minipage}{.49\textwidth}
\begin{figure}[H]
\includegraphics[scale=0.3]{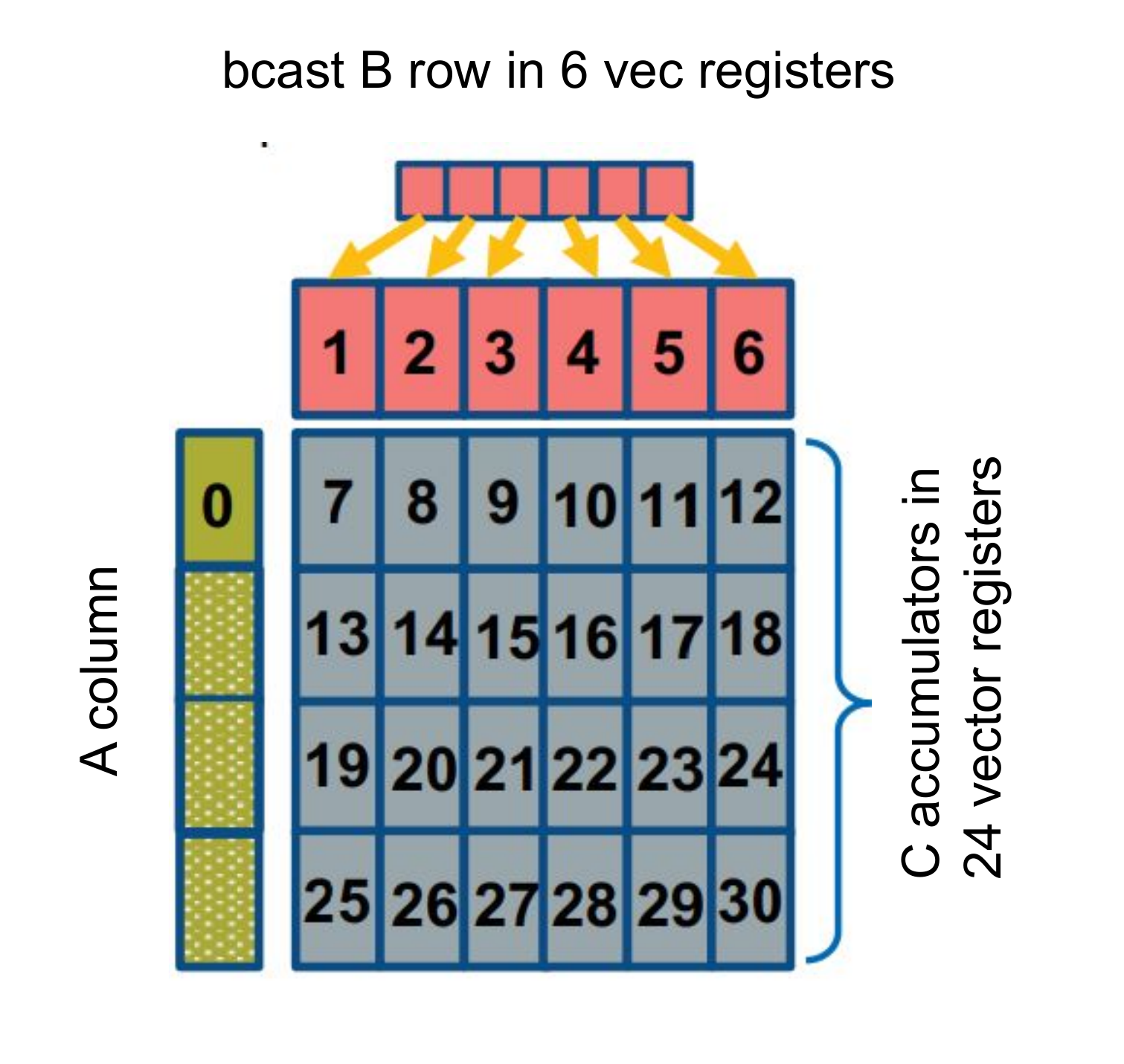}
\caption{Outer product small GEMM microkernel. Source: Georganas et al. \cite{georganas2020harnessing}}
\label{fig:outerproduct}
\end{figure}
\end{minipage}

Algorithm \ref{alg:gemmmicro} shows how the GEMM microkernel is implemented.
The C matrix is brought into registers. The entire matrix multiplication is realized
as a series of outer products: columns of A are multiplied with rows of B.
While the outer products are being computed, either columns of A are reused or rows of B are reused depending on the matrix sizes. During the entire computation,
the C matrix is held in registers and the result of outer products are accumulated
in the C matrix.

Figure \ref{fig:outerproduct} depicts how an example outer product is performed.
In this illustrative example, $m = 64$ and $n = 6$.
Let us consider that there are 32 vector registers
in the underlying architecture and each vector register holds up to 16 tensor elements.
Vector registers 6 through 30 hold the output C matrix.
($m \times n = 64 \times 6 = 384$ elements of the C matrix can fit in 24 vector registers, namely, 7 - 30.)
A row of B is broadcast into 6 vector registers. 
The first 16 elements of the first column of matrix A are loaded into a register
and multiplied with B's vector registers 1 - 6 via fused-multiply-add (FMA) operations and the accumulator registers 7 - 12 are updated.
The rest of the 48 elements of the first column of matrix A are brought into registers
in a similar fashion and accumulators 13 - 30 are updated.
At this point a column of A would have been completely multiplied with a row of B
and the results would have been accumulated for matrix C in registers 7 - 30.
Later, all $k$ columns of A are streamed in and multiplied with the same row of B.
After the results are accumulated in this fashion, the second row of B would be broadcast to vector registers 1 - 6 and matrix A is streamed in again.
In this set up, arrays B, and C are in registers throughout the whole computation while
matrix A is streamed in $k$ times. 
Finally, the C values in accumulators are stored back to memory. 

We note that this is one of the strategies adopted in LIBXSMM.  
Depending on the $m$ and $n$ values, and the architecture at hand (vector length,
the number of vector registers) different schemes (such as streaming in the B matrix
and completely reusing the A matrix) could be used.
Additionally, software prefetches for matrices A and B are also issued to 
mitigate the cache miss latency overheads.
For a given set of problem sizes, the LIBXSMM framework JIT-compiles the microkernel
(JIT compilation - Just In Time compilation). It emits assembly instructions and provides a function pointer to the JITed microkernel.

\subsection{Evaluation of Compile Time Selection of Top Performing Code Version}

We evaluate the efficacy of the developed techniques on two prominent DL operators, viz., convolutions and GEMMs.

\subsubsection{Convolutions}
\label{experiments:convs}
We use the PolyDL system to optimize the convolutions of Resnet-50 \cite{he2016deep}, Fast R-CNN (\textsf{fastrcnn}) \cite{girshick2015fast},  Mask R-CNN (\textsf{maskrcnn})  \cite{DBLP:journals/corr/HeGDG17},
 the popular
and the state-of-the-art image recognition neural network models.
We also measure the performance of the same convolutions using 
the implementations from the Intel oneDNN library 
and those obtained via auto-tuning with the AutoTVM system.
We pick the top 1 variant
the code generator produces, i.e., $k = 1$. That is, a single version is selected.

\begin{minipage}{0.7\textwidth}
\begin{figure}[H]
\begin{lstlisting}[language=C,basicstyle=\tiny,frame=bottomline]
#pragma omp parallel for private(ofm_tile, ifm_tile, ij, oj, kj, ki, ii)
for (img = 0; img < nImg; ++img) {
 for (ofm_tile = 0; ofm_tile < nOfm / GEMM_BLOCK; ++ofm_tile) {
  for (ifm_tile = 0; ifm_tile < nIfm / GEMM_BLOCK; ++ifm_tile) {
   for (oj = 0; oj < ofh; ++oj) {
	ij = oj * STRIDE_H;
	for (kj = 0; kj < kh; ++kj) {
	 for (ki = 0; ki < kw; ++ki) {

	  /* GEMM operation begins */
	  for (oi = 0; oi < ofw; ++oi) {
	   ii = oi * STRIDE_W;
	    for (ofm = 0; ofm < GEMM_BLOCK; ++ofm) {
		for (ifm = 0; ifm < GEMM_BLOCK; ++ifm) {
		 output[img][ofm_tile][oj][oi][ofm] +=
		  filter[ofm_tile][ifm_tile][kj][ki][ifm][ofm] 
		   * input[img][ifm_tile][ij+kj][ii+ki][ifm];
		}
	   }
	  }
	 /* GEMM operation ends */
     }}}}}}
\end{lstlisting}
\caption{The 2-D Convolution  code}
\label{fig:convcode}
\end{figure}
\end{minipage}
\begin{minipage}{0.3\textwidth}
Figure \ref{fig:convcode} shows the convolution code.
The shown code is data tiled in the input and output channel 
dimensions.
The convolution code has a matrix multiplication operation
(denoted \emph{GEMM} in the code) embedded in it.
We use the performance obtained using the code
shown in \ref{fig:convcode} as the \emph{baseline}.
The \emph{GEMM} (matrix multiplication) operation in the Figure will be replaced with a call to the LIBXSMM implementation of matrix multiplication.
\end{minipage}

\emph{PolyDL} performs outer loop optimization around the call to the matrix multiplication microkernel
by loop reordering and tiling using various tile sizes.
We show the performance obtained by inserting
the LIBXSMM microkernel in the code listed in Figure \ref{fig:convcode}
under the banner of \textsf{Microkernel} in the subsequent performance
graphs.
Comparing the performance of \textsf{Microkernel} with 
\textsf{PolyDL} will show the need to perform outer loop tuning
as done by \textsf{PolyDL} to obtain high performance for all layers
and for all models.
Depending on the tensor sizes, we generate different number of code variants for each layer.
The number of variants generated varies from 5 to 21.
This is because, the number of tile sizes we can explore is a function of the tensor sizes.
We generate a larger number of variants for convolutions on larger tensors and fewer variants for convolutions on smaller tensors.
On average, for each layer, 11 versions are generated. Consequently, the task of the PolyDL system is to rank the generated variants based on performance.
Each program is run a 1000 times and the average performance across those runs
is reported in the paper.

The machine has a 512-bit SIMD vector unit and supports
AVX-512 vector instructions. 
Consequently, 16 floating point arithmetic operations can be performed
at a time (each floating point number is 32 bits long, and therefore,
16 floating point numbers make up 512 bits: $32 \times 16 = 512$).
Since the microkernel vectorizes along the input and 
output channel loops ($ifm$ and $ofm$ loops in the code),
to fully utilize the vector unit, the input and output channel
widths have to be 16 or multiples of 16.
In the CNN models considered, $86\%$ of the convolutions meet
this criterion and those convolutions are selected for experimental
evaluation.
The peak single precision floating point performance of a 28-core
Cascade Lake processor is  \textasciitilde 3,300 GFLOPS/s.
We set the mini-batch size to 28 and use data parallelism:
the convolution operator is applied on 28 images simultaneously.

To train a DNN model for performing ranking of code variants as described
in \S \ref{sec:dnnranking}, we use 70\% of the experimental data collected
(to avoid overfitting) -- that is, performance data of 70\% of the code versions generated are used to form the training data set.
We create a single DNN model using data from all CNN models
and use it to rank variants across the CNN models.

\begin{minipage}{0.7\textwidth}
\begin{figure}[H]
\centering
\includegraphics[scale=0.5]{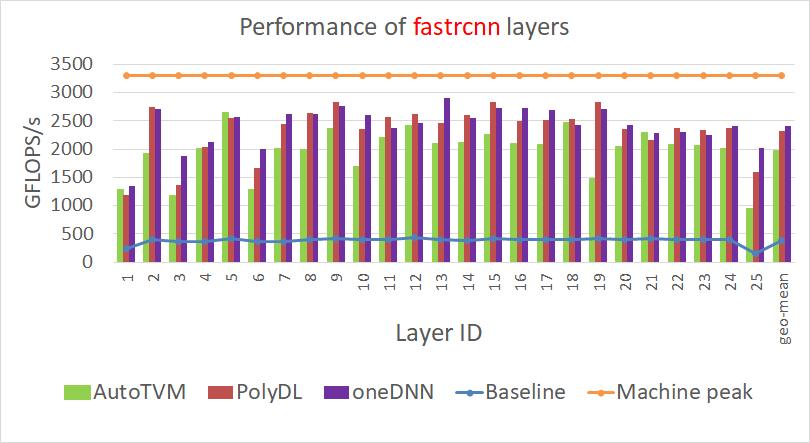}
\caption{Performance of fastrcnn layers}
\label{fig:fastrcnn}
\end{figure}
\end{minipage}
\begin{minipage}{0.3\textwidth}
Figure \ref{fig:fastrcnn} shows the performance in terms of GFLOPS/s
(Giga Floating point Operations per second) of 
the baseline code, PolyDL, AutoTVM and oneDNN 
on convolutions of \textsf{fastrcnn}. 
The PolyDL performance shown is the performance of 
the top code variant selected using the cost modeling based
poly-ranking algorithm described in \S \ref{sec:polyranking}.
\end{minipage}
The performance of \textsf{PolyDL} vis-a-vis the baseline code is 
anywhere between 4X and 11X across layers.
The higher performance of PolyDL is due to 
1) the optimization of outer loops 
2)  the use of optimized GEMM microkernel for the inner loops.
PolyDL performance is close to oneDNN's.
For some layers such as layer 11, PolyDL is 9\% faster than oneDNN
while for a few layers notably layer 1, and 3, oneDNN performs better.
AutoTVM's performance often lags that of PolyDL's.
The geometric average of GFLOPS/s numbers are also shown
in the graph. They are 1965, 2322, and 2408 for AutoTVM, PolyDL, and oneDNN respectively.

\begin{minipage}{0.7\textwidth}
\begin{figure}[H]
\centering
\includegraphics[scale=0.5]{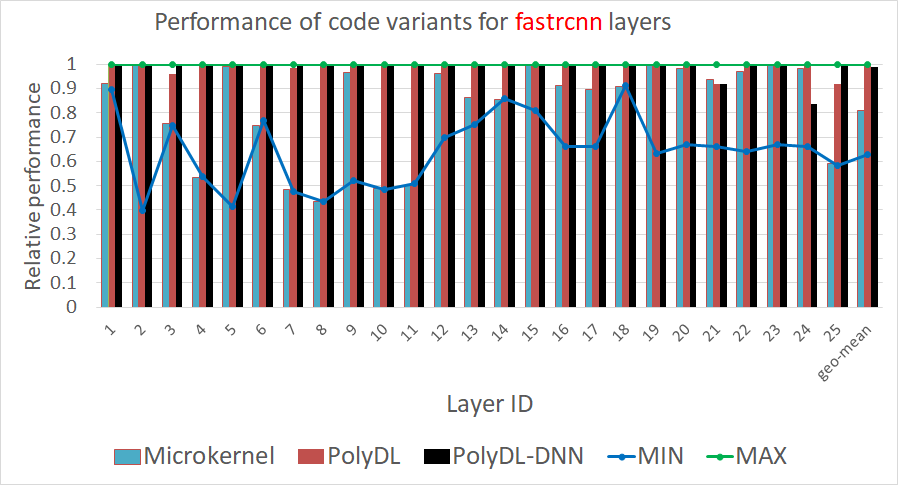}
\caption{Performance distribution of code variants}
\label{fig:fastrcnn_distro}
\end{figure}
\end{minipage}
\begin{minipage}{0.3\textwidth}
Figure \ref{fig:fastrcnn_distro} shows the performance distribution for all
layers of \textsf{fastrcnn}. 
The performance is normalized with respect to that of the best
performing variant found empirically.
\end{minipage}
The crux of the PolyDL technology presented in the paper
is to rank a given set of code variants 
using compile-time static analysis.
Therefore, the closer the performance of the PolyDL picked
version is to the maximum performance seen by any code 
variant explored, the more efficacious the PolyDL algorithms are.
In the graph we show the minimum performance observed,
the maximum performance seen, the performance of the variant with default loop order shown in Figure \ref{fig:convcode}
with \emph{microkernel} inserted -- \textsf{Microkernel},
the performance of the code
picked per the poly-ranking algorithm (\S \ref{sec:polyranking})
-- \textsf{PolyDL}
and the performance of  the code picked per the DNN based ranking algorithm
(\S \ref{sec:dnnranking}) -- \textsf{PolyDL-DNN}.
Here, we see that the performance distribution is great:
the difference between the performance of the best and the worst code variant 
seen is vast for all layers except layers 1, and 18. We observe that PolyDL is
able to pick a variant whose performance is close to the performance of the 
best performing version.
\emph{In the case of \textsf{fastrcnn}, we see that \textsf{PolyDL}
outperforms \textsf{Microkernel} significantly clearly showing the need
for outer loop tuning in addition to having a high performance implementation 
of matrix multiplication in the inner most loops.}
\textsf{PolyDL} picked code achieves \textasciitilde 2X performance gains over 
the code with the default loop order for layers 4, 7, 8, 10, and 11
while for layer 25, \textsf{PolyDL} is 56\% higher performing.
Across all layers of \textsf{fastrcnn}, \textsf{PolyDL} improves
the performance over the default loop order by 28\% on average.


\begin{minipage}{0.7\textwidth}
\begin{figure}[H]
\centering
\includegraphics[scale=0.5]{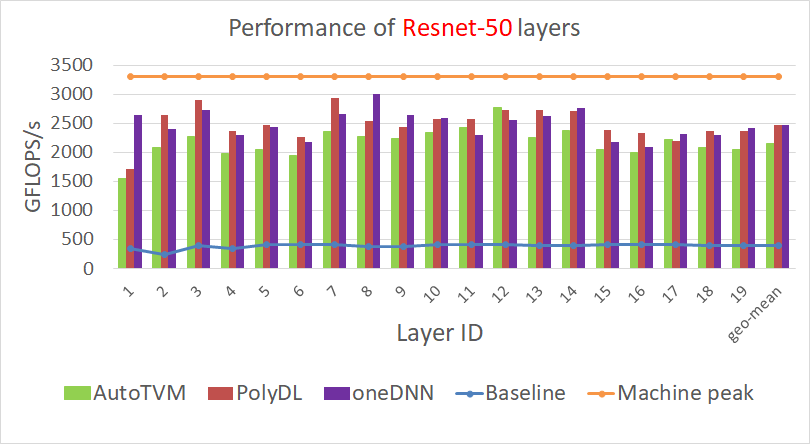}
\caption{Performance of Resnet-50 layers}
\label{fig:resnet}
\end{figure}
\end{minipage}
\begin{minipage}{0.3\textwidth}
The performances achieved by different methods for the convolutions of 
\textsf{resnet} are shown in Figure \ref{fig:resnet}.
The performance of PolyDL over the baseline is 
5X to 10X for all layers.
In most cases, PolyDL closely matches the performance of
oneDNN library.
\end{minipage}
In several instances, PolyDL outperforms oneDNN, 
notably for layers with IDs 7, 11, 15, and 16 where
the performance gain is over 10\%.
On some layers such as layer 1, oneDNN fares better.
This is explained by customizations for specific problem sizes including
insertion of careful data prefetching instructions in the
 oneDNN library code.
In contrast, PolyDL's approach is automatic and 
in the case of Resnet-50, we observe that we are able to attain
the same performance levels as oneDNN overall.
PolyDL is 14\% higher performing than AutoTVM on average.
\begin{minipage}{0.7\textwidth}
\begin{figure}[H]
\centering
\includegraphics[scale=0.45]{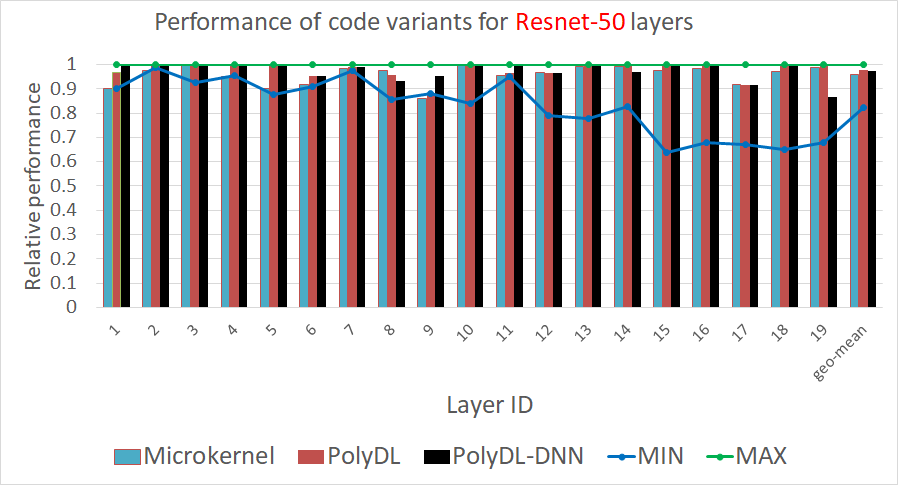}
\caption{Performance distribution of code variants}
\label{fig:resnet_distro}
\end{figure}
\end{minipage}
\begin{minipage}{0.3\textwidth}
Figure \ref{fig:resnet_distro} shows the performance 
distribution of code variants generated for each layer of Resnet-50.
We note that the performance of the \textsf{PolyDL} version
is close to the maximum performance in most layers save layer 9.
\end{minipage}
Even though in terms of cache behavior (PolyDL primarily models the cache
behavior), the variant selected by PolyDL may be the best, other factors such as
prefetching, TLB behavior etc may cause its performance to be lower than those
of other variants. 
The minimum performance seen i.e., the performance of the worst 
code variant, varies across layers -- for layer 12 through 19,
the minimum performance is much farther from the maximum
performance. For the initial layers however, the different
code variants generated perform similarly.
For Resnet-50, performance levels of \textsf{Microkernel} and \textsf{PolyDL} are similar indicating that the original loop order shown in 
Figure \ref{fig:convcode} gets good performance.
Even so, for layer 1, \textsf{PolyDL} is 7\% higher performing than \textsf{Microkernel}.
We observe that there is not a considerable difference
in the performance achieved by 
the cost model based ranking method -- \textsf{PolyDL}, 
and the DNN
based ranking method -- \textsf{PolyDL-DNN}.

In Figure \ref{fig:maskrcnn} and Figure \ref{fig:maskrcnn_distro}, 
we show the performance
achieved by various systems and the performance distribution of code variants
seen for the CNN model -- \textsf{maskrcnn}.

\begin{minipage}{0.7\textwidth}
\begin{figure}[H]
\centering
\includegraphics[scale=0.5]{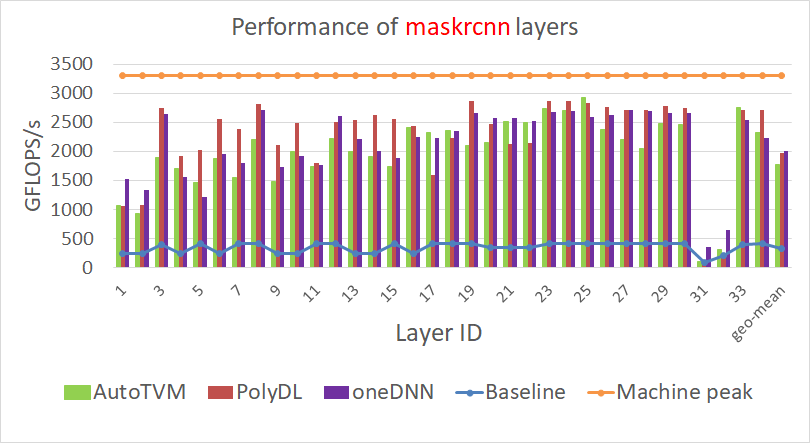}
\caption{Performance of maskrcnn layers}
\label{fig:maskrcnn}
\end{figure}
\end{minipage}
\begin{minipage}{0.3\textwidth}
In Figure \ref{fig:maskrcnn} we observe that the performances of two layers
of \textsf{maskrcnn}
-- layer 31, and 32 are very low compared to the machine peak.
The reason is, the image sizes for the two layers are 7X7 and 1X1 respectively.
\end{minipage}
Consequently, the amount of work that each core has to perform is less and therefore,
all three systems -- AutoTVM, PolyDL, and oneDNN are not able to attain performance close to the machine
peak.

\begin{minipage}{0.7\textwidth}
\begin{figure}[H]
\centering
\includegraphics[scale=0.45]{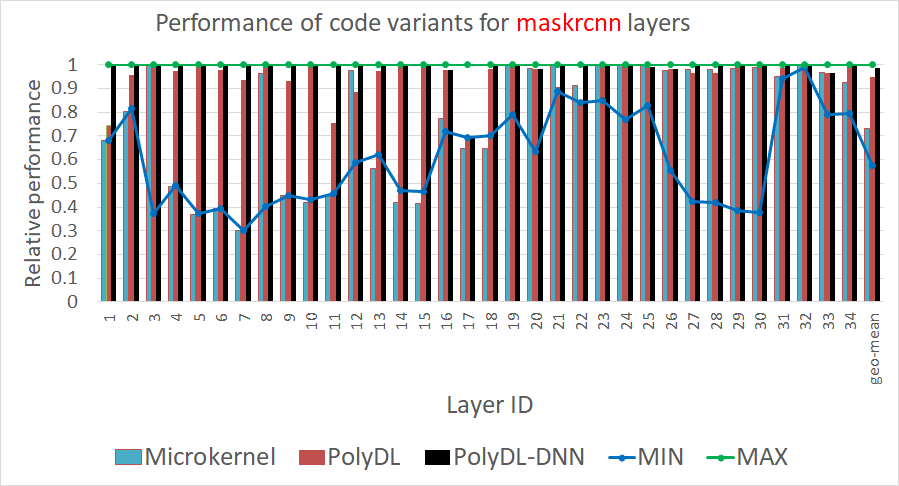}
\caption{Performance distribution of code variants}
\label{fig:maskrcnn_distro}
\end{figure}
\end{minipage}
\begin{minipage}{0.3\textwidth}
For \textsf{maskrcnn} too, we discover that the default loop order -- \textsf{Microkernel}, leaves a lot of performance on the table:
for layers 4, 5, 6, 9, 10, 14, 15, \textsf{PolyDL} gets more than 2X extra 
performance compared to only the use of the microkernel. 
For layer 7, \textsf{PolyDL} is 3X higher performing than \textsf{Microkernel}.
\end{minipage}
Across all layers of \textsf{maskrcnn}, on average \textsf{PolyDL} is 1.29X faster compared to \textsf{Microkernel}.

Across different models,
the performance of \textsf{PolyDL-DNN} is consistently
slightly better than that of \textsf{PolyDL}. 
Further, \textsf{PolyDL} achieves magnitudes of higher performance
compared to the baseline code and 
is very competitive with respect to the hand crafted oneDNN library code.
AutoTVM generates and runs typically over 1000 code variants for each layer and takes $\sim$15 - 20 minutes to discover the
best performing variant for a single layer. The PolyDL method zeroes in on the best version in under one minute.
We note that AutoTVM's methodology is not fully automatic. For example, we have obtained the performance results using CPU X86 specific implementations of Convolution code
within AutoTVM \cite{autotvmx86guide}, \cite{autotvmx86code}.
AutoTVM developers have created CPU specific and additionally architecture specific (i.e., Cascade Lake) code within the framework.
Although AutoTVM discovers good tile sizes through auto-tuning, a lot of customized code has also been written to obtain high performance
on Cascade Lake CPUs. 
Therefore, for the purposes of this experimental evaluation, AutoTVM represents a combination of library development and auto-tuning approach.

\textbf{Comparison with prior compiler works:}
In this paper, we presented an approach to characterizing the working set sizes of the loops
and relating them to the cache sizes of a computer system.
Such an analysis formed the basis for ranking different code variants and selecting the best performing program version in our work.
Alternately, one could compute the number of cache misses for a given program variant 
at different levels of the cache hierarchy and using the number of cache misses approximate its execution time.
Subsequently, the code variant with the smallest estimated execution time can be considered to be the code variant that achieves the highest performance.
Prior works have developed analytical cache miss computation methods \cite{ghosh1997cache,bao2017analytical,gysi2019fast}.
However, none of them can analyze parallel programs. Our PolyDL compiler algorithms presented in this paper can analyze parallel code in addition to sequential code.
Even so, we compare PolyDL with the latest analytical cache miss calculation work, viz., Gysi et al.'s cache modeling work \cite{gysi2019fast} in the following manner. 
In our convolution related experiments, the image -- \textsf{img} loop is parallel (Figure \ref{fig:convcode}).
For the sake of this experimental evaluation, we convert the problem to a sequential one by
assuming that the loop length of the \textsf{img} loop is 1 and each core of the processor gets an equal share of the shared L3 cache.
This is a reasonable assumption because in our experiments, each core processes an image each.
The execution time of a program is estimated to be:
$ L_1~\text{misses} \times \text{lat}^{L_2} + L_2~\text{misses} \times \text{lat}^{L_3} + L_3~\text{misses} \times \text{lat}^{\text{mem}} $.

The running time is equal to the sum of latencies at different cache levels.
One of the summands, for example is, the number of misses at L1 cache is multiplied by the latency of the L2 cache (because L1 misses are serviced from L2 cache).
For this comparison study, we invoke \emph{sequential} analysis in PolyDL too with identical assumptions (\textsf{img} loop length being set to 1, and each processor getting
an equal share of the L3 cache).

\begin{minipage}{0.49\textwidth}
\begin{figure}[H]
\centering
\includegraphics[scale=0.35]{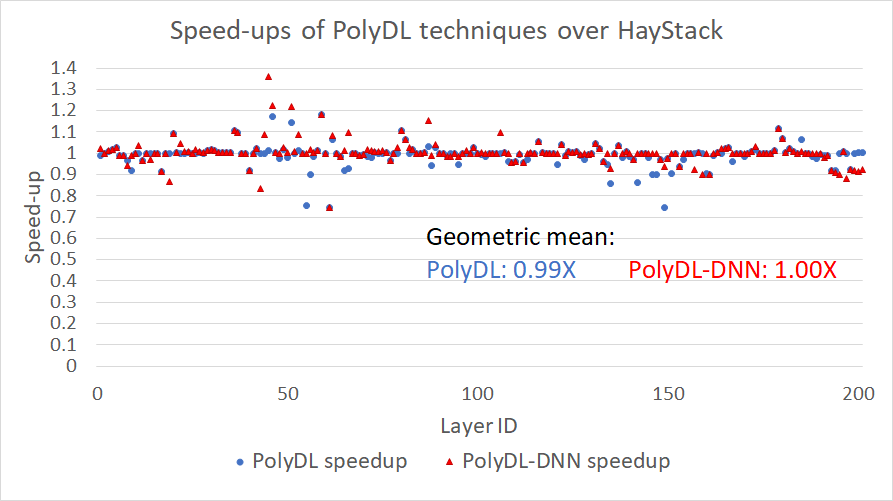}
\caption{Comparison of performance achieved by the code version selected by PolyDL techniques with that of HayStack picked version}
\label{fig:haystackcompare}
\end{figure}
\end{minipage}
\begin{minipage}{0.49\textwidth}
Figure \ref{fig:haystackcompare} shows the speed-ups obtained by \textsf{PolyDL} and \textsf{PolyDL-DNN} with respect to 
Gysi et al.'s cache modeling work \cite{gysi2019fast}. Their tool is termed \textsf{HayStack} and the same name is used in Figure \ref{fig:haystackcompare}.
For most of the total 201 convolution layers (corresponding to the layers of \textsf{fastrcnn}, \textsf{resnet}, \textsf{maskrcnn}, \textsf{xception}, 
\textsf{yolov2}, \textsf{mobilenet}, \textsf{alexnet}, \textsf{overfeat}, \textsf{googlenetv1}, \textsf{googlenetv3}),
 the performances of code versions picked by \textsf{PolyDL}, and \textsf{HayStack} are identical. In some instances \textsf{HayStack} is better
and in others \textsf{PolyDL}/\textsf{PolyDL-DNN} are higher performing. 
\end{minipage}

On average, \textsf{PolyDL-DNN} achieves marginally better performance -- it has a 1.002X speed-up over \textsf{HayStack}.
Between \textsf{PolyDL} and \textsf{HayStack}, \textsf{HayStack} has a slight edge: \textsf{PolyDL} is at 0.99X performance levels of \textsf{HayStack}.
We attribute the observed performance levels of the different methods to the nature of statistics we obtain using the PolyDL algorithms (working set sizes) and those that we obtain using
the HayStack algorithm (the number of cache misses).
It is straight-forward to relate the number of cache misses to the running time of the program using the cache latencies as done above. 
However, relating working set sizes to the running time of the program could be more complicated and the use of DNN techniques (i.e., \textsf{PolyDL-DNN})
is therefore superior to  approximating the running time using cache latencies directly (i.e., \textsf{PolyDL}).

We observed that the running time of the \textsf{HayStack} tool is highly variable. In our experiments, \textsf{HayStack}
took anywhere from a couple of seconds to 37 minutes to process a single code variant. On average, it takes $\sim$ 35 seconds. 
\textsf{PolyDL}'s running time was more uniform -- it takes 2 - 3 seconds for any variant.
Thus, \textsf{PolyDL}'s analysis is orders of magnitude faster than \textsf{HayStack}'s.

\subsubsection{GEMMs} 
\label{experiments:GEMM}
The GEMM operation is at the heart of deep learning \cite{gemmdl,qin2020sigma}.
The fully connected (FC) layers and multi-layer perceptrons map to GEMM operations:
During the forward pass (for inference and for training too), the inputs and the weights are the two matrices that are multiplied.
During the backward pass (for training), the error gradient with respect to the inputs
and the weight matrices are multiplied.

\begin{table}[t]
\caption{\label{tab:GEMMsizes} GEMM sizes}
\small
\begin{tabular}{|c|r|r|r||c|r|r|r|}
\hline
Workload & M & N & K & Workload & M & N & K \\ \hline
GNMT (Machine translation) & 128 & 2048 & 4096 & Synthetic & 4096 & 4096 & 4096 \\ \hline
GNMT (Machine translation) & 320 & 3072 & 4096 & Synthetic & 1024 & 1024 & 32768 \\ \hline
GNMT (Machine translation) & 2048 & 4096 & 32 & Synthetic & 1024 & 32768 & 1024 \\ \hline
DeepBench (General workload) & 1024 & 16 & 500000 & Synthetic &  32768 & 1024 & 1024 \\ \hline
\end{tabular}
\end{table}

We evaluate PolyDL's efficacy on several matrix sizes. The GEMM operation is: $C = \beta . C + \alpha . A . B$ where C matrix's dimensions are 
$M \times N$ while A and B matrices are of size $M \times K$ and $K \times N$ 
respectively. Table \ref{tab:GEMMsizes} lists the GEMM sizes evaluated.
We use several problem sizes  from DL workloads.
We use four synthetic ones also to  
measure the performance  obtained when all M, N, and K sizes are similar
and when one of them is much higher than others (tall-skinny matrices, short-stout ones). We measure the performance of GEMMs when utilizing 32 processor cores.

Figure \ref{fig:baselineGEMM} shows the \emph{baseline} GEMM code. We report the performance obtained by this code when compiled with the icc compiler and ``-O3'' flag in Figure \ref{fig:gemm} -- designated as \textsf{Baseline} in the Figure.
PolyDL framework generates several code variants for GEMM and ranks them.
In Figure \ref{fig:PolyDLGEMM}, we are showing one of the code variants generated.
It is a two level tiled code. Tile sizes -- \textsf{M2\_Tile}, \textsf{N2\_Tile},
\textsf{K2\_Tile}, \textsf{M1\_Tile}, \textsf{N1\_Tile}, and \textsf{K1\_Tile}
are tunable parameters: PolyDL will vary these values and create distinct code variants.
In the inner-most loops the GEMM \emph{microkernel} is invoked for performing
GEMM operation on matrix sizes \textsf{M1\_Tile}, \textsf{N1\_Tile}, and \textsf{K1\_Tile}. Additionally, either loop \textsf{it2} or \textsf{jt2} could be made parallel.
The PolyDL ranking algorithm will chose which loop is to be made parallel.
The number of variants generated varies between 4 (for GEMM sizes M=1024,N=16,K=500000)
and 1228 (for GEMM sizes M=4096,N=4096,K=4096).
On average 491 code variants are explored for each set of GEMM sizes.
We select the top 5\% variants after PolyDL ranks them, and report
the maximum performance obtained among those top 5\% variants.

\begin{minipage}{.49\textwidth}
\begin{figure}[H]
\begin{lstlisting}[language=C,basicstyle=\footnotesize,frame=bottomline]
#pragma omp parallel for private(j, k)
for (i = 0; i < M; i++)
 for (j = 0; j < N; j++)
  for (k = 0; k < K; k++)
  C[i][j] = beta * C[i][j] + 
         alpha * A[i][k] * B[k][j];
\end{lstlisting}
\caption{Baseline GEMM code}
\label{fig:baselineGEMM}
\end{figure}
\end{minipage}
\begin{minipage}{.49\textwidth}
\begin{figure}[H]
\begin{lstlisting}[language=C,basicstyle=\tiny,frame=bottomline]
<@\textcolor{blue}{// First level of tiling}@>
<@\textcolor{magenta}{// Potential parallel loop1: it2}@>
for (it2 = 0; it2 < M; it2 += M2_Tile) {
<@\textcolor{magenta}{ // Potential parallel loop2: jt2}@>
 for (jt2 = 0; jt2 < N; jt2 += N2_Tile) {
  for (kt2 = 0; kt2 < K; kt2 += K2_Tile) {
  <@\textcolor{blue}{// Second level of tiling}@>
   for(it1=it2;it1<it2+M2_Tile;it1+=M1_Tile){
    for(jt1=jt2;jt1<jt2+N2_Tile;jt1+=N1_Tile){
     for(kt1=kt2;kt1<kt2+K2_Tile;kt1+=K1_Tile){
      <@\textcolor{red}{     //Call to GEMM microkernel of size:}@>
      <@\textcolor{red}{     //M1\_Tile,N1\_Tile,K1\_Tile}@>
      microkernel(..)	
      }}}}}}
\end{lstlisting}
\caption{A code version generated for PolyDL}
\label{fig:PolyDLGEMM}
\end{figure}
\end{minipage}
 
We compare the performance obtained by PolyDL with three other systems':
1) Pluto \cite{uday08pldi}, a polyhedral source-to-source compiler,
2) AutoTVM, and 3) oneDNN.
We input the code shown in Figure \ref{fig:baselineGEMM} to the Pluto tool
and obtain tiled and parallelized code. We auto-tune for tile sizes: Pluto 
expects tile sizes to be inputted and we input several tile sizes. 
We report the maximum performance obtained 
by any of the tile size combinations in Pluto.
To obtain the best performance possible from AutoTVM, we follow the GEMM 
optimization recipe provided on the TVM web pages \cite{autotvmgemmcpu}
and additionally explore many tile sizes and report the maximum performance obtained
by any set of tile sizes.

\begin{minipage}{0.7\textwidth}
\begin{figure}[H]
\centering
\includegraphics[scale=0.50]{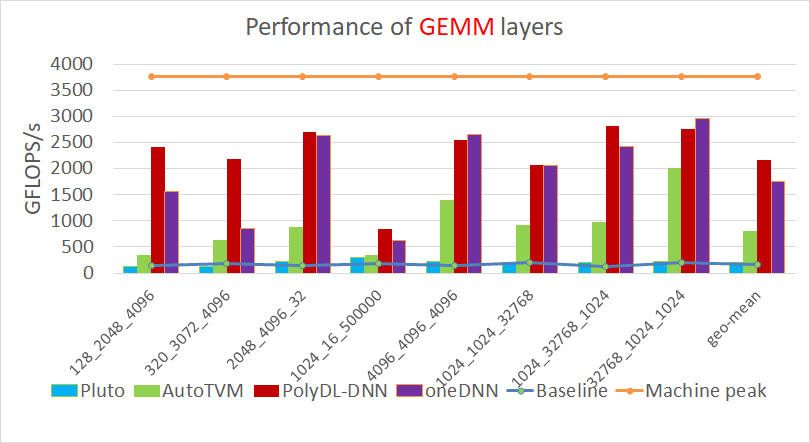}
\caption{Performance of GEMMs for different problem sizes}
\label{fig:gemm}
\end{figure}
\end{minipage}
\begin{minipage}{0.3\textwidth}
Figure \ref{fig:gemm} depicts the performance obtained for various matrix sizes
by different systems. In the graph, on the X-axis, the GEMM sizes (M, N, and K values as M\_N\_K)
are shown. 
We observe that PolyDL-DNN and oneDNN performances
are much higher than the other two, viz., Pluto and AutoTVM.
\end{minipage}
In 5 out of 8 GEMM sizes, PolyDL-DNN obtains substantially higher performance
than oneDNN: For M=320 N=3072 K=4096 sizes, PolyDL-DNN is 2.6X higher performing
than oneDNN. On average, PolyDL-DNN implementations are 1.24X faster compared
to oneDNN's (geometric average).
 PolyDL-DNN GEMM's are 13.4X higher performing than baseline implementations. Pluto performance hovers around the same levels as that of the baseline code. On average, PolyDL-DNN achieves 11.0X speed-up over Pluto.
We note that AutoTVM's performance is orders of magnitude higher than baseline's.
This is partly explained by the fact that explicit vectorization encoded in the 
execution schedule (the most beneficial vectorization loop was already set manually)
which resulted in good vectorization. PolyDL-DNN's GEMMs were 2.7X faster
compared to AutoTVM's on average. 
The average (geo-mean) performance obtained in terms of GFLOPS/s across eight GEMM sizes by different systems, namely,
Baseline, Pluto, AutoTVM, PolyDL-DNN, and oneDNN are 162, 197,
802, 2171, and 1745 respectively. We note that PolyDL achieves the highest performance
among the cohort evaluated in this set of experiments.

In Figure \ref{fig:gemm_distro}, we show the performance spread among
the different code variants that PolyDL system considered. The performances of 
all the code variants are normalized with respect to the maximum performance
recorded for any variant. \textsf{MIN} denotes
the lowest performance of any variant witnessed.
Performances achieved by PolyDL-DNN (ranking of variants using the DNN model described in \S \ref{sec:dnnranking})
and PolyDL (ranking of variants using the heuristics based on cache latencies as presented in \S \ref{sec:polyranking})
are shown. The closer the PolyDL's and PolyDL-DNN's performances are to 1.0 (the MAX line),
the better it is -- it shows that they are able to rank the variants correctly.
We observe that performances of PolyDL-DNN picked code variants are closer to the maximum performance -- within 0.95X of maximum performance.
This shows that through compile-time modeling of data reuses in the loops as described in the paper,
our techniques are able to identify high performance configurations for the loops.

\begin{minipage}{0.6\textwidth}
\begin{figure}[H]
\centering
\includegraphics[scale=0.35]{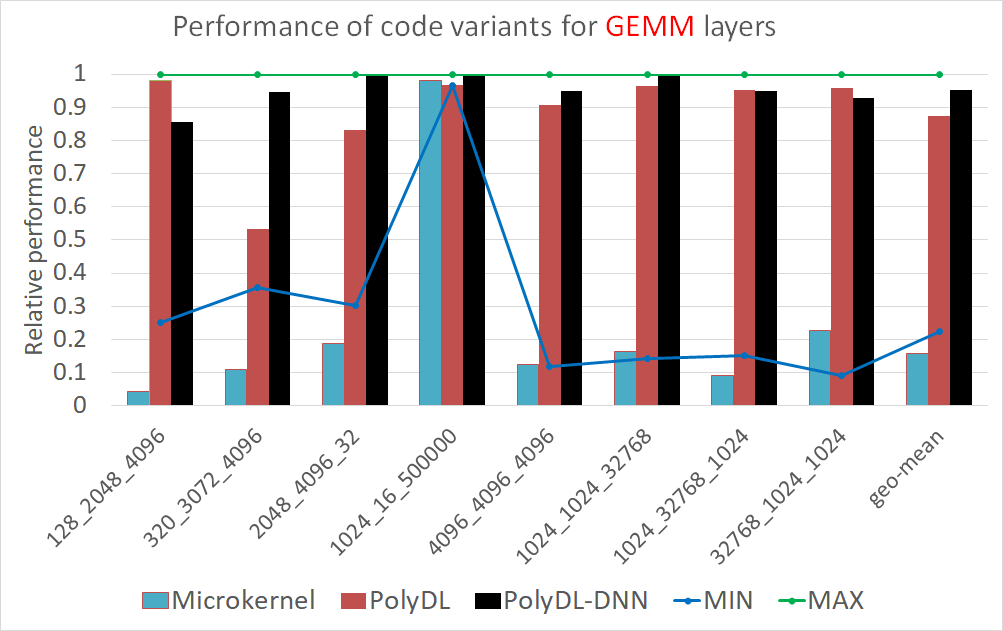}
\caption{Performance distribution of code variants}
\label{fig:gemm_distro}
\end{figure}
\end{minipage}  
\begin{minipage}{0.4\textwidth}
\emph{We also show the performance obtained when the GEMM microkernels are used without
any outer loop tuning in the graph -- denoted as \textsf{Microkernel}}.
To measure the performance by using microkernels alone, we tile the baseline
code shown in Figure \ref{fig:baselineGEMM} by factors of 16 in the three loop dimensions, and invoke the GEMM microkernel to perform matrix multiplications on matrices of sizes 16X16.
\end{minipage}

The microkernel-performance alone is quite low. Nevertheless, 
the use of microkernels alone speeds up baseline GEMM implementations by 2.2X.
PolyDL-DNN achieves 6.0X speed-up over \textsf{Microkernel} implementations.
This underscores the importance of performing outer loop optimizations to obtain 
high performance.

Between PolyDL and PolyDL-DNN, PolyDL-DNN picked code variants have higher performance, on average by a factor of 1.09X. For cost model based ranking of code variants, we
use memory hierarchy's latency and bandwidth values. While it fetches good rankings,
DNN model is able to learn the latency and bandwidth values more accurately
and can learn underlying hardware characteristics such as data prefetching automatically. Therefore, PolyDL-DNN's rankings tend to be slightly better.

%

\subsection{Evaluation of Operator Fusion}
\begin{figure*}[h!]
\centering
\begin{minipage}{0.49\textwidth}
\centering
\includegraphics[scale=0.4]{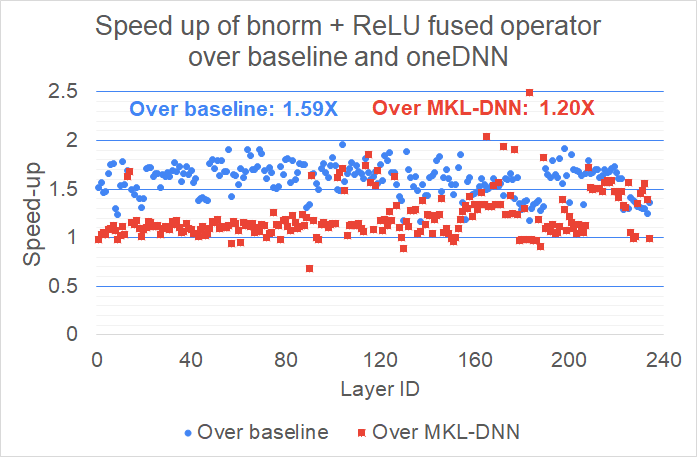}
\caption{Speed up achieved by bnorm and ReLU fused operator over that of unfused baseline code and oneDNN implementation (higher the better).}
\label{fig:bnormfuse}
\end{minipage}
\begin{minipage}{0.49\textwidth}
\centering
\includegraphics[scale=0.4]{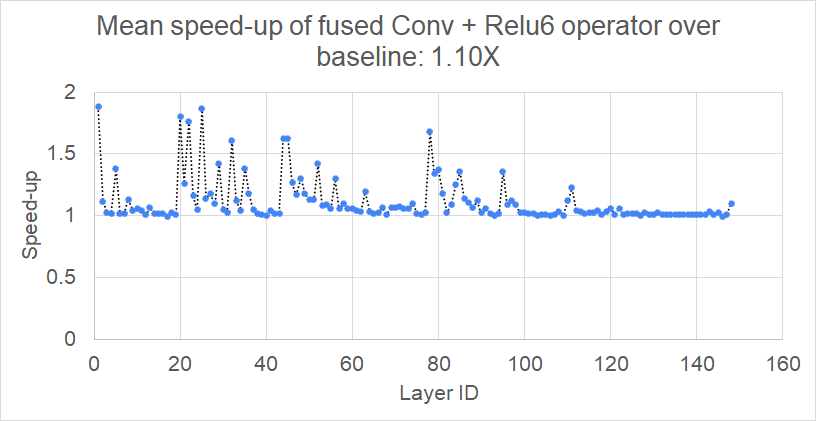}
\caption{Speed up achieved by conv and ReLU6 fused operator over that of unfused baseline code (higher the better).}
\label{fig:convrelu6fuse}
\end{minipage}
\end{figure*}

We evaluate the benefits of operator fusion algorithm presented in the paper ( \S\ref{sec:fusion}) on two 
sequences: 1) batch normalization followed by the activation function ReLU,
2) convolution followed by the activation function ReLU6, which is a variant of ReLU.
The activation function ReLU is defined as: $y = \max(x, 0)$, while ReLU6 is defined as
$y = \min(\max(x, 0), 6)$ \cite{krizhevsky2010convolutional}, that is, the output value is capped at 6. 
The batch normalization (\textsf{bnorm} for short) and ReLU sequence occurs frequently in CNN models including that of Resnet-50. 
We compare the performances of 1) bnorm, ReLU unfused code which forms the baseline 
2) fused bnorm + ReLU operator using our fusion algorithm 3) fused bnorm + ReLU operator in the oneDNN library. 
There is no facility in AutoTVM to perform operator fusion automatically and therefore, we do not compare its performance on the sequence.
Figure \ref{fig:bnormfuse} shows the speed-ups obtained by our fused operator vis-a-vis the baseline code and oneDNN's corresponding fused operator for the 
tensor sizes of all layers of CNN models we have considered in our experiments.
The speed-ups achieved by the fused operator are 1.59X and 1.20X (geometric average across tensor sizes) compared to the unfused baseline code and the oneDNN's operator respectively.
Batch normalization is a memory bandwidth bound operation and therefore, fusing the subsequent ReLU operation reduces round trips to the main memory which substantially
improves the performance.
We perform fusion experiments with the convolution and ReLU6 sequence too. 
We compare the performances of unfused and fused versions of the sequence.
We note that ReLU6 is not supported in the oneDNN library. 
Researchers have found that ReLU6 improves the performance of image recognition models \cite{krizhevsky2010convolutional}.
However, ReLU6 is not widely adopted and it is not supported in the oneDNN library  presumably because the high investment in supporting it is not justified.
This underscores that 1)  hand coding of a plethora of DNN primitives researchers experiment with is not scalable, 
2) when an operator is not available in a library like oneDNN, it hampers the data scientists' ability to quickly experiment and refine the DNN models.
Therefore, automatic compilation techniques like the one developed in this paper are needed to address these bottlenecks.

Figure \ref{fig:convrelu6fuse} shows the speed-ups of the fused operator when compared to the corresponding unfused operator. 
The ReLU6 activation function is applied on the output of the convolution.
When the size of the output tensor is large, we observe a higher speed-up. When the output size is smaller,  fusion has a marginal benefit.
On average, the fused operator is 1.10X faster (geometric average). Furthermore, convolution is a compute-bound operation and therefore, 
the time spent in it is large relative to the time spent in the ReLU6 operation. 
This explains the contrasting speed-ups seen in the two sequences we have evaluated. 
For the bnorm + ReLU sequence, the speed-ups are larger because one, bnorm is a memory bound operation and two, the time spent in it
is less compared to convolution. Consequently, the time spent in ReLU as a percentage of the time spent in bnorm is larger.
Therefore, fusing the two operators shows larger performance gains for the bnorm + ReLU sequence.
\section{Related Work}
\label{sec:related}
We discuss related works from polyhedral compilation, auto-tuning systems,
deep learning DSL (Domain Specific Language) frameworks, and GEMM microkernel research.

\textbf{Polyhedral compilation techniques} have been developed for source-to-source code
transformation for better cache locality and parallelization.
The Pluto  compiler \cite{uday08pldi} derives an execution schedule
for the code that attempts to minimize data reuse distances.
The effect of the Pluto transformation will be that 
the iterations that use the same data will be executed close to each other in time
and therefore, it will be cache friendly.
The Pluto algorithm can accomplish fusion of loops too.
However, Pluto's 
performance can be far from what we can achieve with the use of microkernels that exploit
the vector hardware effectively and by doing outer loop tuning
in the way we have developed this work: In Section \ref{experiments:GEMM},
we show that \emph{our techniques can produce on average 11.0X higher performing GEMM 
implementations}.
Furthermore, the Pluto algorithm experiences scalability issues -- 
as the number of loops in a loop nest increases, the Pluto algorithm/tool can take
exceedingly long time to derive a schedule.
When we inputted the convolution followed by ReLU code sequence, the tool took several
hours and did not produce an output.
Kong et al. \cite{Kong:2019:MTM:3314221.3314653} develop a framework to decompose a program into sub-parts and use customized criteria (as opposed to using a single objective function) -- such as stride optimization, outer loop optimization, inner loop optimization to transform code. They show that their work without tiling transformation is able to achieve comparable results to that of Pluto. 

The existing polyhedral compilation techniques fail to achieve performance competitive with hand tuned libraries. The main reason is, the loop transformations polyhedral
techniques encode operate at a high level (source-to-source) and therefore,
are unable to perform low-level orchestration of vector register allocation, detailed instruction scheduling (e.g., like that of the GEMM microkernel implementation described in \S \ref{subsection:GEMMmicrokernel}). The latter aspects are crucial to achieving high performance on CPUs.
An effective approach we believe is to use the polyhedral model based loop scheduling for outer loops for efficient use of cache hierarchy and to use microkernels for effective vectorization in the inner loops such as the one we have proposed in this paper.

Bondhugula et al. \cite{bondhugula2010model} propose a \textbf{fusion model} that considers loss of parallelism with aggressive fusion. The number of available
hardware prefetch streams is also used as a constraint to determine the beneficial
fusion structures. In this paper, we have proposed a domain specific fusion algorithm
that fuses heavy operators with element-wise operators. We have simplified 
the criteria for fusion while preserving parallelism.
The patterns that our algorithm tackles occur frequently in DL workloads and our algorithm presents a first generic approach towards automatic fusions of operators in 
 deep learning.
Cornwall et al. \cite{cornwall2007explicit} discuss using software engineering concepts like component based programming in the context of the development of a visual effects library. They use algorithmic skeletons 
to extract the iteration space for loops, and high level metadata for dependence analysis such that various
optimizations can be applied. They propose using shifting as an enabler for fusion. Along with support for
array-contraction which enables vectorization, the presented techniques result in good performance.


\textbf{Compile-time modeling of cache behavior} and in particular calculating the number of cache misses
has been an active area of research \cite{ghosh1997cache,bao2017analytical,gysi2019fast}.
Researchers have demonstrated good accuracy in predicting the number of cache misses
on simulators. The modern computer architectures employ a hash based scheme to map
memory addresses to cache sets \cite{yarom2015mapping} which breaks
the assumptions behind the cache miss analyses. 
In the present work, we model the behavior of caches as well. 
However, we do not model cache misses rather we consider data reuses and determine
the size of cache needed to exploit the data reuses under conservative conditions.
We ignore streaming accesses as their misses in cache will not be crucial in the resulting
performance. 
Our analysis works on parallel code, whereas prior works are not equipped to handle
parallel loops.
Because of the these improvements, we show that we are able to accurately
rank code variants in terms of performance.
Our experimental evaluation comparing against the latest cache miss analysis work \cite{gysi2019fast}
using sequential loop setting (\S\ref{experiments:convs}) shows that the working set size approach we have adopted
in our present work is as good as using cache misses for modeling of performance or is
slightly better. 


\textbf{DSLs and Autotuning systems:}
\emph{Iterative compilation} \cite{pouchet2008iterative} and/or combined model-driven and iterative compilation techniques have been explored for program optimization \cite{pouchet2010combined}.
TVM \cite{chen2018tvm}, a compiler for deep learning, introduces the concept of \emph{tensorization}, where 
a unit of computation can be replaced with a microkernel written using hardware
intrinsics. 
AutoTVM \cite{chen2018learning} which is based on TVM, is targeted at accelerating deep learning workloads and uses machine learning to guide auto-tuning of deep learning primitives. 
We compared our techniques with AutoTVM in this paper and showed that the DL primitives created by
PolyDL (our work) enjoy superior performance vis-a-vis AutoTVM. Further, the time
it takes for our techniques to create high performance code is a fraction of the time
AutoTVM takes.
Tiramisu \cite{baghdadi2019tiramisu}  is a polyhedral model based compiler framework that introduces a scheduling language to allow the programmer to explore various
program transformations. 
%
Halide \cite{ragan2012decoupling} is a Domain Specific Language (DSL) and framework that introduces the distinction between an \emph{algorithm} and its associated \emph{schedule}.
The DSL provides high level abstractions for the programmer to encode different schedules and thereby, explore different schedules in a productive way and discover 
high performance schedules. 
Adams et al \cite{adams2019learning} automate the task of finding optimal schedules in Halide
using machine learning.
In our present work, we take a compiler-centric approach:
Using compiler-generated features we identify promising schedules (either through 
a heuristic based cost model or with the use of a neural network model).
Because of these contrasting methodologies, Adams' system would require a lot more
training data (and therefore, resources to optimize a given code) than our PolyDL system.
SWIRL \cite{venkat2019swirl} is a system similar to Halide in the sense SWIRL
also allows the programmer to specify the \emph{algorithm}
separately from its \emph{schedule}. It allows one to encode numerous \emph{transformation recipes}. Our PolyDL system can be complementary to 
SWIRL in that, the techniques developed in our work could be used to automate
the task of deriving the beneficial \emph{transformation recipes} automatically.
\textsf{TensorComprehensions} \cite{vasilache2018tensor} and  \textsf{Diesel} \cite{elango2018diesel} are two DSL systems developed for generating efficient code for GPUs. Both systems use the polyhedral model for code generation and provide auto-tuning capabilities. Unlike our system which is geared towards CPUs, both 
\textsf{TensorComprehensions} and \textsf{Diesel} target only GPUs.
Since the considerations for optimizing code for GPUs are quite different
from that for CPUs, their techniques are not directly applicable for CPUs.


\textbf{GEMM microkernel optimizations:}
Goto and van de Geijn \cite{goto2008anatomy} focus on creating high-performance GEMMs by using a layered decomposition over the three micro-kernels, from which the others could be derived. The results show how these three lowest level decompositions can achieve high performance, thus
resulting in an overall high performance for GEMM implementations.
Springer et al. \cite{springer2018design} use Tensor contraction
for reducing the rank of matrices, followed by packing the tensor operands of the macro-kernel in the available
caches, thus attaining the high performance.
To remove the need of customization of microkernel for each hardware Veras et al. \cite{veras2016automating} present an
automated approach by decomposing the microkernel (GEMM) into unit updates and building different algorithms over it.
All these works focusing on \emph{microkernels} are complementary to our approach:
we could leverage microkernels created from any of these frameworks for our inner loops,
thereby increasing the \emph{programmer productivity} in our PolyDL approach.

\textbf{Two level optimization of code:}
Barthou et al \cite{barthou2007loop} present a hierarchical compilation model, wherein
the outer loops are optimized for data locality, while the inner loops are abstracted 
into kernels and are optimized for ILP (Instruction Level Parallelism) using the backend compiler. While at a high level,
their approach and ours are similar, there are some crucial differences:
1) The outer loop optimization as developed in our work is more sophisticated (working set size enumeration and subsequent use of a DNN model for ranking of code variants)
compared to the selection of tile sizes such that the data accessed in a tile fit in a certain level of cache as adopted
in Barthou et al's work. 2) The reliance on the backend compiler for kernel optimization
in their work can lead to substantially lower performance compared to what is achievable
through the use of \emph{microkernels}. E.g., In our experimental evaluation (\S \ref{experiments:GEMM}), we find that AutoTVM's (their approach is nearly similar to AutoTVM's) performance is substantially lower than PolyDL's. 
3) Their techniques are not applicable for parallel code and consequently, for multi-core
architectures while handling of parallel code is baked into our techniques.

The BLIS  framework \cite{zee2016blis} shows that BLAS routines can be instantiated
using a single DGEMM microkernel. The microkernel needs to be customized for a given hardware and the programmer has to tune outer loops for cache use by choosing
appropriate block (tile) sizes for the target architecture.
Our PolyDL approach is similar to BLIS's in spirit: we decompose the problem of optimizing an operator into two subproblems -- one, performing outer loop optimization,
and use of a microkernel for the inner loops. In our framework, we use novel polyhedral
model based cache data reuse algorithms to perform outer loop optimizations automatically. Additionally, the use of a neural network based approach, and operator fusion algorithms are new and are beneficial  for deep learning workloads. 

\section{Conclusion}
\label{sec:conclusion}
In this paper, we presented novel compiler algorithms 
to derive high performing DL primitive implementations automatically
and to perform operator fusions.
We proposed a methodology to optionally use \emph{microkernels} for the inner most loops of DL primitives to optimally use
the vector pipelines of modern CPUs. 
With a combination of the above techniques, we demonstrated through
experimental evaluation that we are able to match the performance of 
expert coded implementations of the Intel oneDNN library for CNNs and GEMMs.
Additionally, because our method works at compile-time, we require much less
time and compute resources to derive efficient implementations compared to 
auto-tuning systems such as AutoTVM.
Our system -- \textsf{PolyDL} will ease the development of computer architecture specific libraries by at most requiring  the development of only a small number of \emph{microkernels}.
Additionally, it will allow data scientists to enjoy high performance on 
the new DNN architectures they develop immediately and automatically, without
waiting for their DNN constructs to be implemented in a library by expert programmers.

\balance
\bibliographystyle{ACM-Reference-Format}
\tiny{
\bibliography{paper}
}


\end{document}